\documentclass[%
twocolumn,
superscriptaddress,
amsmath,amssymb,
aps,
pra,
floatfix,
nofootinbib
]{revtex4-1}

\usepackage{graphicx}
\usepackage{dcolumn}
\usepackage{bm}
\usepackage{grffile} 
\usepackage{subfigure}
\usepackage[marginal]{footmisc}
\usepackage{epstopdf}
\usepackage{amssymb}
\usepackage{mathrsfs}
\usepackage{amsmath,mathtools,amsfonts,bm,graphicx}
\usepackage{upgreek}
\usepackage{color}
\usepackage{etoolbox}
\usepackage[colorlinks=true,linkcolor=blue,urlcolor=blue,citecolor=blue]{hyperref} 
\usepackage{multirow}
\usepackage[normalem]{ulem}
\usepackage{scrextend}

\begin{document}

\preprint{APS/123-QED}
\title{Exact Thermodynamics For Weakly Interacting Normal-Phase Quantum Gases: Equations of State For All Partial Waves}

\author{Xin-Yuan Gao}
\affiliation{%
Department of Physics, The Chinese University of Hong Kong, Shatin, New Territories, Hong Kong, China
}%
\author{D. Blume}
\affiliation{%
Homer L. Dodge Department of Physics and Astronomy, The University of Oklahoma, 440 W. Brooks Street, Norman, Oklahoma 73019, USA
}%
\affiliation{
Center for Quantum Research and Technology, The University of Oklahoma, 440 W. Brooks Street, Norman, Oklahoma 73019, USA
}
\author{Yangqian Yan}%
 \email{yqyan@cuhk.edu.hk}
\affiliation{%
Department of Physics, The Chinese University of Hong Kong, Shatin, New Territories, Hong Kong, China
}
\affiliation{
The Chinese University of Hong Kong Shenzhen Research Institute, 518057 Shenzhen, China
}%

\date{\today}

\begin{abstract}
While the thermodynamics for bosonic systems with weak $s$-wave interactions has been known for decades, a general and systematic extension to higher partial waves has not yet been reported. 
We provide closed-form expressions for the equations of state for weakly interacting systems with arbitrary partial waves in the normal phase. 
Thermodynamics, including contact, loss rate, and compressibility, are derived over the entire temperature regime.
Our results offer an improved thermometer for ultracold atoms and molecules with weak high-partial wave interactions. 
\end{abstract}

\maketitle

\section{Introduction} The equations of state (EOS) for {\em{ideal non-interacting}} Bose and Fermi gases are standard textbook results~\cite{huang2008statistical}
that are of immense importance to cold atom experiments. For example, temperatures of weakly interacting quantum gases are frequently extracted by fitting experimental data to non-interacting density profiles. While weak interactions modify the non-interacting density profile only slightly, recent molecular quantum gas experiments~\cite{bohn2017cold,anderegg2018laser,ni2008high, park2015ultracold, seesselberg2018modeling, takekoshi2014ultracold, molony2014creation, guo2016creation, rvachov2017longlived,duda2023longlived} suggest that the chemical reaction rate is comparatively sensitive to the interactions even in the weak-interaction limit.  The reason is that the contact~\cite{tan2008energetics,tan2008generalized,tan2008large,stewart2010verification,kuhnle2010universal,sagi2012measurementa,werner2012generala,wild2012measurements,hoinka2013precise,shashi2014radiofrequency,werner2012generalb,yu2015universal,luciuk2016evidence,yoshida2015universal,he2016concept,zhang2017contact,song2020evidence}, which is the thermodynamic variable that governs the chemical rate in the weakly-interacting regime~\cite{braaten2008exact,braaten2013universal,braaten2017lindblad,he2020universal,gao2023temperaturedependent}, changes from zero for non-interacting systems to a finite value for interacting systems.

This article is devoted to the EOS of single-component Bose and Fermi gases with {\em{weak}} interactions in the normal phase. The EOS is well understood and available in an analytical form for single-species bosons with weak $s$-wave interactions~\cite{andersen2004theory}. In contrast, for single-component Fermi gases with weak $p$-wave interactions, the contacts and EOS have only been studied in the low- and high-temperature regimes~\cite{ding2019fermiliquid,he2020universal,gao2023temperaturedependent}, 
even though higher partial-wave physics has attracted increased attention recently~\cite{yao2019degenerate,cui2017observation,liu2018feshbach,shi2023collective}.
Analytical expressions for the EOS of single-component gases beyond the $s$-wave case (i.e., for $p$-wave Fermi gases, $d$-wave Bose gases, $f$-wave Fermi gases, etc.)---applicable over the entire temperature regime---do not exist. 

Within the Hartree-Fock framework, we derive analytical closed-form expressions for the EOS, applicable to all temperatures, of single-component atomic or molecular quantum gases with weak $l$-wave interactions in the normal phase.
We calculate the contact, which determines the chemical reaction rate of ultracold gases. Using the virial expansion, we find that, while the contact of weakly-interacting $s$-wave Bose gases in the normal phase is a pure two-body quantity, that of weakly-interacting $p$-wave Fermi gases displays pronounced three-body effects even at temperatures as high as the degeneracy temperature. 
This effect is shown to arise from many-body dressing, i.e., the emergence of quasi-particles at leading order in the interaction strength.
We also discuss the relation between the resulting reaction rate and that obtained through a simple thermal average over the inelastic cross-section. 
Applying the local-density approximation (LDA), we calculate the contacts of the harmonically trapped systems. Our results show that the trapped system needs to be cooled to rather low temperatures to probe the ``low-temperature" portion of the EOS of the homogeneous system.

The article is arranged as follows: Section~\ref{SecII} introduces the $l$-wave low-energy two-body interaction potential employed in Sec.~\ref{SecIII} to derive the normal-phase $l$-wave EOS in the weak-interaction limit. Section~\ref{SecIV} applies the EOS of the homogeneous system to deduce explicit, yet general, expressions for the two-body contact and two-body loss rate coefficient, which are interpreted using the virial expansion (see Sec.~\ref{SecV}). Section~\ref{SecVI} focuses on the homogeneous $l=0$ and $l=1$ systems. The loss rate coefficients derived in this work are compared with heuristic thermal averages in Sec.~\ref{SecVII}. Section~\ref{SecVIII} applies the homogeneous EOS to harmonically trapped systems using the LDA. Finally, Sec.~\ref{SecIX} discusses the applicability regime of the theory results derived in this work, while Sec.~\ref{SecX} concludes. Technical details are relegated to several appendices.

\section{Interaction Model} 
\label{SecII}
The low-energy two-body potential for arbitrary partial-wave channel $l$ reads
\begin{equation}
    U_l(\mathbf{q},\mathbf{q}')=4\pi g_l q^l (q')^l\sum_{m=-l}^l {Y}_{lm}(\hat{\mathbf{q}}){Y}^*_{lm}(\hat{\mathbf{q}}'),
    \label{interaction}
\end{equation}
where ${Y}_{lm}(\hat{\mathbf{q}})$ is the spherical harmonic, and $\mathbf{q}$ and $\mathbf{q}'$ are the incoming and outgoing relative momenta~\cite{ho2005fermion}. The two-body phase shifts $\delta_l$ are given by $k^{2l+1}\cot(\delta_l)=-1/a_l+\mathcal{O}(k^2)$, where $a_l$ is the scattering length in the $l$th partial-wave channel.
To describe the binding energy of shallow two-body bound states, the leading-order effective-range correction needs to be included, and $\delta_l$ needs to be expanded up to order $k^2$~\cite{landau1977quantum}. However, effective-range corrections can be excluded since we work in a weakly interacting regime where bound states do not contribute. A standard renormalization procedure gives (see Appendix~\ref{appendix:A})
\begin{equation}
    \frac{1}{g_l}=\frac{M}{4\pi\hbar^2a_l}+\frac{M}{2\pi^2\hbar^2}\int_{0}^{\infty}dqq^{2l}.
    \label{renormalization_condition}
\end{equation}
Noticing that since we are using first-order perturbation theory where ultraviolet divergencies are absent, renormalization is not required, implying that the bare coupling $g_l$ and scattering length $a_l$ are related by $g_l={4\pi\hbar^2a_l}/{M}$, where $M$ denotes the mass of the gas constituents (atoms or molecules). For $l=0$, the interaction $U_0$ is, as expected, equal to $g_0$~\cite{fetter2012quantum}.
The next section uses the interaction potential 
$U_l(\mathbf{q},\mathbf{q}')$ to derive perturbative results for the normal-phase EOS.

\section{Equations of State in Normal Phase}
\label{SecIII}
To include the two-body interactions in the EOS, we account for the mean-field corrections to the Bose-Einstein distribution function ($l$ even) and Fermi-Dirac distribution function ($l$ odd) in momentum space~\cite{fetter2012quantum},
\begin{equation}
    n_\mathbf{k}=\left[\exp\left(\dfrac{\epsilon_k^{(0)}+\hbar\Sigma_l(\mathbf{k})}{k_B T}\right)z^{-1}\mp1\right]^{-1},
    \label{selfconsistent_nk}
\end{equation}
where $\epsilon_k^{(0)}=\hbar^2k^2/2M$ denotes the single-particle kinetic energy, $\hbar\Sigma_l(\mathbf{k})$ the self-energy, and $\hbar\mathbf{k}$ the momentum. In Eq.~(\ref{selfconsistent_nk}) and in what follows, the upper sign is for even $l$ (single-component bosons) and the lower sign for odd $l$ (single-component fermions). The self-energy reads
\begin{equation}
    \Sigma_l(\mathbf{k})=\frac{2}{\hbar}(2\pi)^{-3}\int d^3k' U_l\left(\frac{\mathbf{k}-\mathbf{k}'}{2},\frac{\mathbf{k}-\mathbf{k}'}{2}\right)n_{\mathbf{k}'},
    \label{HF_approx}
\end{equation}
from which we can obtain the normal-phase grand potential $\Omega$ by the ``generalized Hellman-Feynman theorem"~\cite{fetter2012quantum} 
\begin{equation}
    \Omega-\Omega^{(0)}=\frac{V}{2}\int_0^1\frac{d\lambda}{\lambda}\int\frac{d^3k}{(2\pi)^3}\hbar\Sigma_l(\mathbf{k},\lambda)n_{\mathbf{k}}(\lambda),
    \label{Hellman-Feynman_theorem}
\end{equation}
where $n_\mathbf{k}(\lambda)$ and $\Sigma(\mathbf{k},\lambda)$ are defined through Eqs.~(\ref{selfconsistent_nk}) and (\ref{HF_approx}) with the two-body potential $U_l(\mathbf{q},\mathbf{q}')$ scaled by $\lambda$~\footnote{Namely,
$
    n_\mathbf{k}(\lambda)=\left[\exp\left(\frac{\epsilon_k^{(0)}+\hbar\Sigma_l(\mathbf{k},\lambda)}{k_B T}\right)z^{-1}\mp1\right]^{-1}
$ and $
    \Sigma_l(\mathbf{k},\lambda)=\frac{2}{\hbar}(2\pi)^{-3}\int d^3k' \lambda U_l\left(\frac{\mathbf{k}-\mathbf{k}'}{2},\frac{\mathbf{k}-\mathbf{k}'}{2}\right)n_{\mathbf{k}'}(\lambda)
$}.
Here,
\begin{equation}
\Omega^{(0)}=\mp k_B T V\frac{\mathrm{Li}_{5/2}(\pm  z)}{\lambda_T^3}
\end{equation}
is the non-interacting grand potential, where
$$
\lambda_T=\hbar\sqrt{\frac{2\pi}{M k_B T}}
$$
is the thermal wavelength, and $z$ is the fugacity. 
At the leading order in the scattering length, we find
\begin{equation}
\begin{split}
    \frac{\Omega}{k_B T V}\approx&\mp\frac{\mathrm{Li}_{\frac{5}{2}}(\pm z)}{\lambda_T^3}+\frac{a_l}{\lambda_T^{2l+4}}\sum_{i,j,n\atop i+j+n=l}\mathcal{C}(i,j,n,l)\\
    &\times\mathrm{Li}_{\frac{2i+n+3}{2}}(\pm z)\mathrm{Li}_{\frac{2j+n+3}{2}}(\pm z),
    \label{Omega}
\end{split}
\end{equation}
where the indices $i,j,$ and $n$ start from $0$, and
\begin{equation}
\begin{split}
    \mathcal{C}(i,j,n,l)=&(2l+1)\pi^{l-1}\frac{(1+(-1)^n)2^{n+2}l!}{i!j!n!(1+n)}\\
    &\times\Gamma\left(\frac{2i+n+3}{2}\right)\Gamma\left(\frac{2j+n+3}{2}\right).
\end{split}
\end{equation}
Here, $\mathrm{Li}_s$ and $\Gamma$ are the polylogarithm and gamma functions, respectively.
To construct the full EOS, the mean particle density $n$ needs to be expressed in terms of $z$. We achieve this by treating the self-energy as a small parameter and integrating Eq.~(\ref{selfconsistent_nk}) in momentum space:
\begin{equation}
\begin{split}
  n=&\pm\frac{\mathrm{Li}_{\frac{3}{2}}(\pm z)}{\lambda_T^3}-\frac{2a_l}{\lambda_T^{2l+4}}\sum_{i,j,n\atop i+j+n=l}\mathcal{C}(i,j,n,l)\\
  &\times\mathrm{Li}_{\frac{2i+n+1}{2}}(\pm z)\mathrm{Li}_{\frac{2j+n+3}{2}}(\pm z).   
\end{split}
\label{n}
\end{equation}
One can check that Eqs.~(\ref{Omega}) and (\ref{n}) fulfill the thermodynamic relation $n=-\frac{1}{V}\frac{\partial \Omega}{\partial \mu}=-z\frac{\partial(\Omega/k_B T V)}{\partial z}$, where $\mu$ denotes the chemical potential. 
Equations~(\ref{Omega}) and (\ref{n}) are the first main result of this article. From Eqs.~(\ref{Omega}) and (\ref{n}), one can---at least formally---calculate all thermodynamic quantities. 
Fully analytical expressions for the isothermal compressibility, entropy, and isochoric heat capacity are given in Appendix~\ref{appendix:B}.

\section{Contact and Two-body Loss}
\label{SecIV}
In addition to the observables considered in Appendix~\ref{appendix:B}, we consider the contact $C_l$, which is conjugate to the inverse scattering length. The contact has been discussed extensively for the two-component Fermi gas at unitary~\cite{stewart2010verification,kuhnle2010universal,sagi2012measurementa,werner2012generala,wild2012measurements,hoinka2013precise,shashi2014radiofrequency}. Working in the grand canonical ensemble, where the fugacity $z$ is a thermodynamic variable, $C_l$ is defined in terms of the grand potential $\Omega$ by the adiabatic relation
\begin{equation}
    \frac{(2l+1)\hbar^2 C_l}{2M}=-\frac{\partial \Omega}{\partial a_l^{-1}}.
    \label{adiabatic_relation}
\end{equation}
This definition of $C_l$ generalizes the definition of the $p$-wave contact $C_1$~\cite{luciuk2016evidence}. For $s$-wave interacting Bose gases, the most commonly employed definition of the contact $C_0$ differs from Eq.~(\ref{adiabatic_relation}) by a factor of $2\pi$~\cite{tan2008energetics,tan2008generalized,tan2008large}. The description of higher-partial wave systems typically requires a second contact, namely the conjugate to the effective range~\cite{yu2015universal,luciuk2016evidence,yoshida2015universal,he2016concept,zhang2017contact}. 
Since we find that it affects the thermodynamics of weakly interacting systems at sub-leading order, we exclude it from our discussion.

The contact of weakly-interacting systems is a fascinating thermodynamic quantity since it determines the loss rate due to chemical reactions between two particles. Examples of reactions in molecular NaRb and KRb gases are: 
$$
\begin{aligned}
&\mathrm{NaRb}+\mathrm{NaRb}\rightarrow\mathrm{Na_2}+\mathrm{Rb_2}~(s\mathrm{-wave~Bose~gas})\\
&\mathrm{KRb}+\mathrm{KRb}\rightarrow\mathrm{K_2}+\mathrm{Rb_2}~(p\mathrm{-wave~Fermi~gas})
\end{aligned}
$$
When the two incoming reactants are ``scattered" into final products, the (typically large)
binding energy is converted to the kinetic energy of the products. Consequently, the
products have so much energy that they are not held in place by the comparatively shallow trapping potential. Since the reaction time is short compared to the typical time scale of experimental observations, a non-hermitian Hamiltonian with a complex interaction potential can effectively describe the process. 
For the single-component $p$-wave gas,
it was shown that the change of the number $N$ of constituents is related to the imaginary part of the scattering length~\cite{gao2023temperaturedependent},
\begin{equation}
    \label{eq_help7}\frac{\mathrm{d}N}{\mathrm{d}t}=\frac{4}{\hbar}\langle\mathrm{Im}(H)\rangle=\frac{4}{\hbar}\left.\frac{\partial \Omega}{\partial a_1}\right|_z\mathrm{Im}(a_1),
\end{equation}
where $H$ is the effective Hamiltonian with complex interaction and $\langle\cdot\rangle$ denotes the thermal average. Since the derivation in Ref.~\cite{gao2023temperaturedependent} was done in
real space, without making any assumptions about the form of the interaction, the result can be straightforwardly generalized to arbitrary partial-wave channels:
\begin{equation}
    \frac{\mathrm{d}N}{\mathrm{d}t}=\frac{4}{\hbar}\left.\frac{\partial \Omega}{\partial a_l}\right|_z\mathrm{Im}(a_l).
\end{equation}
From Eq.~(\ref{eq_help7}) and the definition of the contact, Eq.~(\ref{adiabatic_relation}), one obtains
\begin{equation}
    \frac{dn}{dt}=2(2l+1)\frac{\hbar}{M}\frac{C_l}{V}\frac{\mathrm{Im}(a_l)}{[\mathrm{Re}(a_l)]^2}=-\beta_l n^2,
    \label{loss_and_contact}
\end{equation}
where $n$ denotes the particle density; the loss-rate coefficient $\beta_l$ can be measured experimentally~\cite{ni2010dipolar,ospelkaus2010quantumstate,demarco2019degenerate}. The loss-rate coefficient characterizes---due to the $n^2$ term---losses that arise from two-body processes. 
In general, though, the loss-rate coefficient may be $n$-dependent, implying that $dn/dt$ may effectively scale with $n^3$ or $n$ to some other power. 

Combining the EOS and the definition of the contact, we find the contact in the canonical ensemble:
\begin{equation}
\begin{split}
    \frac{C_l(z)}{V}=&\frac{4\pi [\mathrm{Re}(a_l)]^2}{(2l+1)\lambda_T^{2l+6}}\sum_{i,j,n\atop i+j+n=l}\mathcal{C}(i,j,n,l)\\
    &\times\mathrm{Li}_{\frac{2i+n+3}{2}}(\pm z^{(0)})\mathrm{Li}_{\frac{2j+n+3}{2}}(\pm z^{(0)}),
\end{split}
\label{general_contact}
\end{equation}
where the fugacity $z^{(0)}$ of the non-interacting system is implicitly determined by $\mathrm{Li}_{3/2}(\pm z^{(0)})=\pm n \lambda_T^3.$ Equation~(\ref{general_contact}) and its interpretation and implications (see below) are the second main result of this paper.

\section{Virial Expansion Analysis}
\label{SecV}

To unravel how the many-body thermodynamics emerges from the two-body scattering length and few-body correlations, we employ the virial expansion, which provides a systematic expansion in terms of one-, two-, three-, and higher-body clusters~\cite{huang2008statistical}. Formally, we expand $\Omega$ in terms of the fugacity $z$, 
\begin{equation}
\Omega=-k_B T Z_1\sum_{j=1}^{\infty}b_jz^j,
\end{equation}
where $Z_1=V/\lambda_T^3$ is the canonical partition function for a single constituent in a box with volume $V$. 
The determination of the virial coefficient $b_j$ requires information up to the canonical partition function $Z_j$ for $j$ constituents~\cite{liu2013virial}, i.e., $b_j$ contains one-, two-, $\cdots$, $j$-body physics; $Z_j$ with $j>1$ accounts for interactions as well as exchange statistics. Since we have an analytical expression for $\Omega$, the virial coefficients $b_j$ can be calculated analytically up to arbitrarily large $j$ by Taylor expanding Eq.~(\ref{Omega}) around $z=0$. We provide expressions for $\Delta b_j=b_j-\frac{(\pm 1)^{j-1}}{j^{5/2}}$ up to $j=4$:
\begin{equation}
\begin{split}
    \Delta b_1&=0,\\
    \Delta b_2&=-\frac{a_l}{\lambda_T^{2l+1}}\sum_{i,j,n\atop i+j+n=l}\mathcal{C}(i,j,n,l),\\
    \Delta b_3&=\mp\frac{a_l}{\lambda_T^{2l+1}}\sum_{i,j,n\atop i+j+n=l}\mathcal{C}(i,j,n,l)2^{-\frac{1+2i+n}{2}},\\
    \Delta b_4&=-\frac{a_l}{\lambda_T^{2l+1}}\sum_{i,j,n\atop i+j+n=l}\mathcal{C}(i,j,n,l)\left[2^{-3-l}+2\times3^{-\frac{3+2i+n}{2}}\right],
\end{split}
\label{virial_coefficients}
\end{equation}
The expressions for $\Delta b_j$ will be interpreted below.
The following section applies the $l$-wave result for $C_l$ to two commonly investigated systems, namely $s$-wave Bose (in this case, our virial coefficients agree with the literature~\cite{marcelino2014virial}) and $p$-wave Fermi gases.

\section{Homogeneous Systems}
\label{SecVI}
\subsection{Single-component \textit{s}-wave Bose gas}
The contact $C_{0}$ for the weakly-interacting single-component Bose gas, applicable to any temperature $T$ above the transition temperature $T_C$, is directly proportional to $n^2$:
\begin{equation}
    \frac{C_{0}}{V}=\frac{8\pi \left[\mathrm{Re}(a_0)\right]^2}{\lambda_T^6}[\mathrm{Li}_{\frac{3}{2}}(z^{(0)})]^2=8\pi \left[\mathrm{Re}(a_0)\right]^2n^2.
    \label{s-wave_contact}
\end{equation}
Since the quantity $n^2$ can be interpreted as the semi-classical pair density, the thermodynamic variable $C_0$ is a two-body quantity in the weak-interaction limit; in other words, many-body dressing is absent. As a consequence, the corresponding loss-rate coefficient $\beta_0$ is independent of $n$, 
\begin{equation}
    \beta_0=-\frac{16\pi\hbar\mathrm{Im}(a_0)}{M}.
\end{equation}
Even above degeneracy,  two-body chemical reactions of the weakly-interacting single-component $s$-wave gas do not exhibit three- or higher-body correlations. This behavior can be traced back to how the self-energy modifies the momentum distribution Eq.~(\ref{selfconsistent_nk}). At the mean-field level, the $s$-wave interactions lead to a self-energy $\Sigma_0$ that is independent of $\mathbf{k}$, i.e., $\Sigma_{0}(\mathbf{k})=\Sigma_{0}$~(see Appendix~\ref{appendix:B1}). According to Eq.~(\ref{selfconsistent_nk}), the interactions can thus be interpreted as modifying the chemical potential without modifying the character of the constituents, i.e., the constituents remain free particles, and each two-body collision involves exactly two ``physical" constituents. The virial expansion formalism can further confirm the interpretation. By self-consistently calculating the contact with truncated virial expansion at $j$, we find that $j=2$ is enough to produce the exact results and that choosing a higher $j$ does not introduce new terms~(see Appendix~\ref{appendix:C}). 

\subsection{Single-component \textit{p}-wave Fermi gas} 
\begin{figure}[t!]
    \centering
    \includegraphics[width=0.49\textwidth]{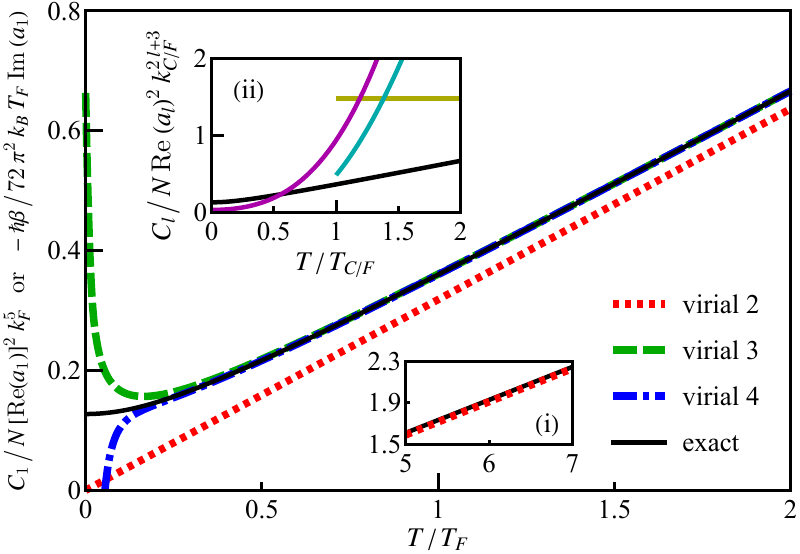}
    \caption{Contact (or two-body loss-rate coefficient), both in scaled dimensionless form, for single-component $p$-wave gas as a function of scaled temperature. The solid line shows Eq.~(\ref{p-wave_contact}); dotted, dashed, and dash-dotted lines show the second-, third-, and forth-virial expansions.  Inset~(i): Extension to larger $T/T_F$, illustrating that the second-order virial expansion converges to the exact result at relatively high temperatures. Inset~(ii): Contact for---from bottom to top at $T=T_{C/F}$---$p$-wave (black), $d$-wave (cyan), $s$-wave (yellow), and $f$-wave (magenta).}
    \label{fig1}
\end{figure}

Setting $l=1$ in Eq.~(\ref{general_contact}), we find
\begin{equation}
    \frac{C_{1}}{V}=-\frac{24\mathrm{Re}(a_1)^2\pi^2}{\lambda_T^5}n\mathrm{Li}_{\frac{5}{2}}(-z^{(0)}).
    \label{p-wave_contact}
\end{equation}
Since the polylogarithm on the right-hand side of Eq.~(\ref{p-wave_contact}) has the index $5/2$ as opposed to $3/2$, the polylogarithm cannot, contrary to the $s$-wave case, be directly converted to $n$. Consequently, $C_1$ features a non-trivial dependence on $n$ and $T$. At high temperatures ($z^{(0)}\rightarrow0$), Eq.~(\ref{p-wave_contact}) becomes 
\begin{equation}
    \frac{C_{l=1}(n)}{V}\xrightarrow{T\rightarrow\infty}-\frac{24[\mathrm{Re}(a_1)]^2\pi^2}{\lambda_T^2}n^2.
\end{equation}
In this regime, the contact $C_1$ has---similar to the contact $C_0$ ---a two-body nature. However, unlike in the $s$-wave case, the high-temperature $p$-wave contact has an explicit temperature dependence. Since $\lambda_T^{-2}$ is directly proportional to $T$, $C_1$ increases linearly with temperature, i.e., reactions become slower as the gas is getting colder. The corresponding $\beta_1$ at high temperatures is independent of $n$ and linearly dependent on $T$, 
\begin{equation}
\beta_1\xrightarrow{T\rightarrow\infty}-\frac{72\pi k_BT\mathrm{Im}(a_1)}{\hbar}.
\end{equation}
In the zero temperature limit, Eq.~(\ref{p-wave_contact}) approaches
\begin{equation}
    \frac{C_{l=1}(n)}{V}\xrightarrow{T\rightarrow0} \frac{12}{5}6^{2/3}\pi^{7/3}[\mathrm{Re}(a_1)]^2n^{8/3}.
\end{equation}
Appendix~\ref{appendix:D} discusses how to evaluate the zero-temperature limits of some of the functions that enter into the $l=1$ EOS.
Since the $n$-dependence deviates from $n^2$, $\beta_1$ is $n$-dependent,  
\begin{equation}
\beta_1\xrightarrow{T\rightarrow0}-\frac{144\pi}{5}\frac{k_BT_F}{\hbar}\mathrm{Im}(a_1),
\end{equation}
where $T_F=\frac{\hbar^2}{2M}(6\pi^2n)^{2/3}$ denotes the Fermi temperature.
The black solid line in Fig.~\ref{fig1} shows Eq.~(\ref{p-wave_contact}) as a function of $T/T_F$. The $T^1$- and $T^0$-scalings in the high- and low-temperature regimes fully agree with previous works~\cite{he2020universal,gao2023temperaturedependent}. 

To interpret the change of the dependence of $C_1$ from being proportional to $n^{2/3}$ at low temperatures to being proportional to $n^2$ at high temperatures, we first note that $p$-wave interacting gases may exist in the normal phase approximately all the way down to zero temperature since the superfluid transition temperature is exponentially small~\cite{ho2005fermion}. It is then natural to assume that many-body effects will modify the reaction rate in the low-temperature limit as the incoming and outgoing momenta are expected to be constrained due to the fermionic exchange statistics, i.e., intuitively, one expects some dressing of the constituents due to many-body effects. 
A careful analysis of the self-energy confirms this picture. 
Substituting $\Sigma_{1}(\mathbf{k})=A_1+B_1k^2$ (see Appendix~\ref{appendix:B1}) into Eq.~(\ref{selfconsistent_nk}), the constant $A_1$ can be shown to modify, just as in the $s$-wave case, the chemical potential. The $B_1k^2$ term, in contrast, modifies the single-particle energies $\epsilon_k^{(0)}$, thereby effectively renormalizing the mass of the physical constituents. When a chemical reaction happens at low temperatures, two quasi-particles with effective mass interact instead of two physical constituents. Since the renormalization of the mass is due to many-body dressing, the chemical reaction involves more than two physical constituents.

The above analysis is complemented by the virial expansion up to the fourth order in $z$. Figure~\ref{fig1} compares the contact $C_1$, calculated up to second, third, and fourth order, with the exact result, Eq.~(\ref{p-wave_contact}).
Figure~\ref{fig1} shows that the second-order expansion agrees with the exact expression at $T \gg T_F$ [see inset~(i)]. Importantly, the second-order virial expansion deviates notably from the exact result for temperatures as high as $T/T_F=2$. 
The third-order virial expansion provides an excellent description down to $T/T_F \approx 0.25$. Interestingly, the fourth-order virial expansion does not yield much improvement over the third-order expansion, indicating that three-body processes are essential in chemical reactions of weakly interacting $p$-wave gases for $T/T_F\approx 0.25-2$. At higher temperatures, three-body processes contribute very little. At lower temperatures, the chemical reactions acquire many-body characteristics.

Extending the analysis to higher partial waves, we find that the $C_l$ for $l>1$ also have non-negligible three-body contributions in the vicinity of the degeneracy temperature ($T_C$ for even $l$ and $T_F$ for odd $l$). In the high-$T$ limit, $C_l$ is directly proportional to $T^l$; this scaling is consistent with the two-particle Bethe-Wigner threshold law~\cite{bethe1935theory,wigner1948behavior,sadeghpour2000collisions}. The inset~(ii) of Fig.~\ref{fig1} plots our analytical expressions for $C_l$ for $l=0-3$ as a function of temperature.

\section{Statistical Average of Inelastic Cross Section}
\label{SecVII}

\begin{table*}[t]
\begin{tabular}{|c|c|c|}
\hline
  & heuristic thermal average [Eq.~(\ref{two-body_betal_general})] & thermodynamics [Eq.~(\ref{thermodynamics_betal})] \\
\hline
$s$ &   $-16\pi\hbar\mathrm{Im}(a_0)/M$       &      $-16\pi\hbar\mathrm{Im}(a_0)/M$\\
\hline
$p$ &   $\dfrac{-144\pi^2\hbar\mathrm{Im}(a_1)\mathrm{Li}_{\frac{5}{2}}(-z^{(0)})}{M\lambda_T^2\mathrm{Li}_{\frac{3}{2}}(-z^{(0)})}$       &     $\dfrac{-144\pi^2\hbar\mathrm{Im}(a_1)\mathrm{Li}_{\frac{5}{2}}(-z^{(0)})}{M\lambda_T^2\mathrm{Li}_{\frac{3}{2}}(-z^{(0)})}$           \\
\hline
$d$ &$\dfrac{-1200\pi^3\hbar\mathrm{Im}(a_2)\mathrm{Li}_{\frac{7}{2}}(z^{(0)})}{M\lambda_T^4\mathrm{Li}_{\frac{3}{2}}(z^{(0)})}$&$\dfrac{-600\pi^3\hbar\mathrm{Im}(a_2)\mathrm{Li}_{\frac{7}{2}}(z^{(0)})}{M\lambda_T^4\mathrm{Li}_{\frac{3}{2}}(z^{(0)})}+\dfrac{-600\pi^3\hbar\mathrm{Im}(a_2)\mathrm{Li}_{\frac{5}{2}}(z^{(0)})^2}{M\lambda_T^4\mathrm{Li}_{\frac{3}{2}}(z^{(0)})^2}$\\
\hline
$f$ &$\dfrac{-11760\pi^4\hbar\mathrm{Im}(a_3)\mathrm{Li}_{\frac{9}{2}}(-z^{(0)})}{M\lambda_T^6\mathrm{Li}_{\frac{3}{2}}(-z^{(0)})}$&$\dfrac{-2940\pi^4\hbar\mathrm{Im}(a_3)\mathrm{Li}_{\frac{9}{2}}(-z^{(0)})}{M\lambda_T^6\mathrm{Li}_{\frac{3}{2}}(-z^{(0)})}+\dfrac{-8820\pi^4\hbar\mathrm{Im}(a_4)\mathrm{Li}_{\frac{5}{2}}(-z^{(0)})\mathrm{Li}_{\frac{7}{2}}(-z^{(0)})}{M\lambda_T^6\mathrm{Li}_{\frac{3}{2}}(-z^{(0)})^2}$\\
\hline
\end{tabular}
\caption{Comparison between results from Eq.~(\ref{two-body_betal_general}) and Eq.~(\ref{thermodynamics_betal}) up to $l=3$.}
\label{tab1}
\end{table*}

In the literature, the loss-rate coefficients $\beta_l$ have been calculated by thermally averaging the two-body inelastic cross sections $\sigma_{\text{in},l}(E)$. 
In what follows, we review the steps taken within this approach to derive $\beta_l$~\cite{quemener2010stronga,jachymski2014quantumdefect,waseem2017twobody}.
According to the definition of the scattering length $a_l$ for the $l$-th partial-wave channel, namely $k^{2l+1}\cot(\delta_l)=-1/a_l$, the scattering matrix element $S_l$ in the low-energy threshold limit reads
\begin{equation}
    S_l=e^{2i\delta_l}\approx (1+2\mathrm{Im}(a_l)k^{2l+1})-2\mathrm{Re}(a_l)k^{2l+1}i.
\end{equation}
The inelastic partial-wave cross-section $\sigma_{\text{in},l}$ is related to the scattering matrix element through~\cite{mott1971theory,balakrishnan1997complex,kurlov2017twobody,croft2020unified}
\begin{equation}
    \sigma_{\mathrm{in},l}=(2l+1)\pi\frac{1-|S_l(k)|^2}{k^2},
\end{equation}
where the factor $2l+1$ originates from the fact that the $l$th partial-wave channel has a $(2l+1)$-fold degeneracy.  Assuming $|\mathrm{Im}(a_l)|k^{2l+1}\ll1$ and $|\mathrm{Re}(a_l)|k^{2l+1}\ll1$,
we find
\begin{equation}
    |S_l|^2\approx(1+2\mathrm{Im}(a_l)k^{2l+1})^2\approx 1+4\mathrm{Im}(a_l)k^{2l+1}.
\end{equation}
Utilizing the definition of the scattering energy $E=\hbar^2k^2/2\mu$, where $\mu=M/2$ is the two-body reduced mass, one obtains 
\begin{equation}
    \sigma_{\mathrm{in},l}(E)=-4\pi(2l+1)\mathrm{Im}(a_l)(M E)^{l-1/2}/\hbar^{2l-1}.
    \label{inelastic_cross_section}
\end{equation}
The loss-rate coefficient $\beta_l$ is then found by
thermally averaging the two-body inelastic cross section $\sigma_{\text{in},l}(E)$ over the Boltzmann distribution function~\cite{quemener2010stronga,jachymski2014quantumdefect,waseem2017twobody}:
\begin{equation}
\begin{split}
    \beta_l &= 2\times\frac{\int_0^\infty dE \sqrt{E} e^{-E/k_B T} \sigma_{\mathrm{in},l}(E) \sqrt{4E/M}}{\int_0^\infty dE \sqrt{E} e^{-E/k_B T}}\\
    &=-2^{5+l}\pi^{1/2+l}\Gamma(3/2+l)\frac{\hbar}{M\lambda_T^{2l}}\mathrm{Im}(a_l),
    \label{inelastic_cross_section_cal}
\end{split}
\end{equation}
where the factor $2$ reflects that one inelastic collision process eliminates two particles. Since Eq.~(\ref{inelastic_cross_section_cal}) employs the Boltzmann distribution function, it is instructive to compare it with the high-temperature limit of the expression for $\beta_l$ derived in this work within the thermodynamic formalism. Our exact result and its high-temperature limit read
\begin{equation}
\begin{split}
    &\beta_l=-\frac{8\pi\hbar}{M \lambda_T^{2l}}\sum_{i,j,n\atop i+j+n=l} \mathcal{C}(i,j,n,l)\\
    &\times\dfrac{\mathrm{Li}_{\frac{2i+n+3}{2}}(\pm z^{(0)})\mathrm{Li}_{\frac{2j+n+3}{2}}(\pm z^{(0)})}{\mathrm{Li}_{\frac{3}{2}}(\pm z^{(0)})\mathrm{Li}_{\frac{3}{2}}(\pm z^{(0)})}\mathrm{Im}(a_l),\\
    &\xrightarrow[]{T\rightarrow\infty} -\frac{8\pi\hbar}{M \lambda_T^{2l}}\left(\sum_{i,j,n\atop i+j+n=l} \mathcal{C}(i,j,n,l) \right)\mathrm{Im}(a_l).
    \label{thermodynamics_betal}
\end{split}
\end{equation}
It can be checked that Eq.~(\ref{thermodynamics_betal}) agrees with Eq.~(\ref{inelastic_cross_section_cal}) for each partial wave channel.
This can be understood because the two particles' center-of-mass and relative momenta obey the Boltzmann distribution separately.

At lower temperatures, however, the thermal average needs to be generally performed over the product of two Bose-Einstein or two Fermi-Dirac distribution functions (the three-body analog is discussed in Ref.~\cite{braaten2008threebody}).
Since the product of two such distribution functions does not, unlike in the case of the Boltzmann distribution function, separate in relative and center-of-mass coordinates, the thermal-average approach does not straightforwardly extend to the low-temperature regime. By naively replacing the classical Boltzmann distribution with the quantum version (Bose-Einstein or Fermi-Dirac distribution), Eq.~(\ref{inelastic_cross_section_cal}) becomes
\begin{equation}
    \beta_l \stackrel{?}{=} 2\times\frac{\int_0^\infty dE \sqrt{E} (e^{E/k_B T}z^{-1}\mp 1)^{-1} \sigma_{\mathrm{in},l}(E) \sqrt{4E/M}}{\int_0^\infty dE \sqrt{E} (e^{E/k_B T}z^{-1}\mp 1)^{-1}}.
    \label{two-body_betal_general}
\end{equation}
The question mark over the equal sign indicates that the expression is not rigorous but instead, deduced heuristically.
The integral in Eq.~(\ref{two-body_betal_general}) can be evaluated analytically, and the results for $l=0-3$ are reported in the second column of Table~\ref{tab1}.
Curiously, a comparison of the thermal-average approach and our exact results (third column of Table~\ref{tab1}) shows that the heuristic thermal-average approach does yield the same expressions for $l=0$ and $l=1$ as the rigorous thermodynamic framework developed in this work.
For the higher partial wave channels ($l=2$ and $l=3$), however, the heuristic approach yields a different temperature dependence. The correction of the heuristic expressions of the loss rate coefficient constitutes the third main result of this paper.

\section{Harmonically Trapped Systems}
\label{SecVIII}
\begin{figure}[t!]
    \centering
    \includegraphics[width=0.49\textwidth]{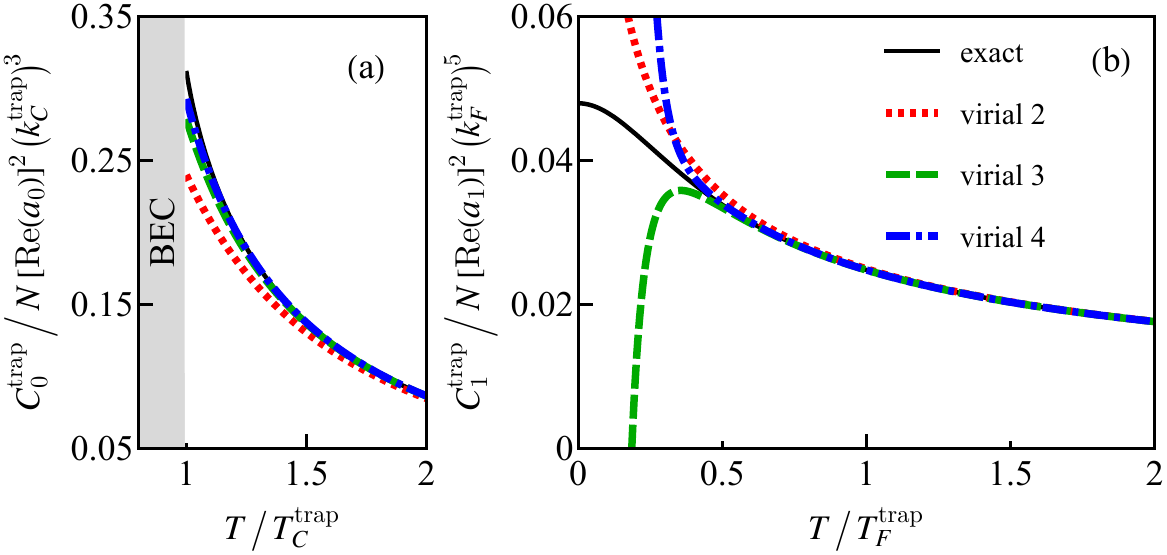}
    \caption{
    Normal-phase contacts, in scaled dimensionless units, for harmonically trapped (a) $s$-wave Bose and (b) $p$-wave Fermi gas. The gray-shaded region in (a) denotes the BEC phase where our calculation does not apply. The exact results (solid lines) are compared with the virial expansions up to fourth order (see legend). Here, $T_C^\mathrm{trap}=(N/\zeta(3))^{1/3}\hbar\omega/k_B$~\cite{dalfovo1999theory}  [$T_F^\mathrm{trap}=(6N)^{1/3}\hbar\omega/k_B$~\cite{butts1997trapped}] is the transition [Fermi] temperature of the non-interacting trapped Bose [Fermi] gas; $\zeta(s)$ denotes the Riemann Zeta function, and $k_C^\mathrm{trap}$ and $k_F^\mathrm{trap}$ denote the momentum scales of the corresponding energy scales.
    }
    \label{fig2}
\end{figure}

We now apply our results to harmonically trapped $N$-particle systems, which are being studied extensively experimentally.
To account for the trap-induced inhomogeneity of the density, we convert our homogeneous EOS, namely Eqs.~(\ref{Omega}) and (\ref{n}), to those for the trapped system via the LDA~\cite{pethick2008bose}. Within this framework, the local density at position $\mathbf{r}$ determines the EOS of the homogeneous system to be used at that point: $\Omega^\mathrm{trap}=\int d^3r \Omega[n(\mathbf{r})]/V.$ To obtain the contact $C_l^{\text{trap}}$ of the trapped system,  Eq.~(\ref{adiabatic_relation}) is evaluated numerically.
The black solid lines in Fig.~\ref{fig2} show the result for a spherically symmetric harmonic trap with angular frequency $\omega$.
To gain physical insight, the EOS of the inhomogeneous system can be described through the virial expansion. In an isotropic harmonically trapped system,
the relation between the virial coefficients $b_j^{\text{trap}}$ of the trapped system and those of the homogeneous system is $b_j^\mathrm{trap}=b_j/j^{3/2}$~\cite{liu2013virial}. The explicit description of the thermodynamics of harmonically trapped single-component gases with weak interactions and the interpretation thereof (see below) constitute the fourth main result of this paper.

Figures~\ref{fig2}(a) and \ref{fig2}(b) compare the contacts of the harmonically trapped $s$-wave Bose and $p$-wave Fermi gas, obtained from the virial expansion up to fourth order, with the full numerical results. We make two observations: (i) For the $s$-wave Bose gas, the contact of the trapped system does not coincide with the second-order virial expansion, indicating that the contact of the trapped system is not, unlike that of the homogeneous system, a two-body quantity. This is because each position in the trap has a distinct local self-energy. Correspondingly, the fugacity of the trapped system cannot be interpreted as a globally shifted fugacity of the non-interacting gas. (ii) The second-order viral expansion for the $p$-wave system works quite well at $T\sim T_F^\mathrm{trap}$;  apparently, three-body corrections play a rather small role near the degeneracy temperature. The two-body (high-temperature) approximation works better for the inhomogeneous system than the homogeneous system since the former is much hotter than the latter for the same scaled temperature (e.g., $T/T_{F}^{\text{trap}}=1$ and $T/T_{F}=1$ correspond to $z \approx 0.17$ and $z=0.98$, respectively). This validates previous works on loss processes of harmonically trapped $p$-wave  systems~\cite{he2020universal,gao2023temperaturedependent}. 
Finally, we note that while the adiabatic relation, Eq.~(\ref{adiabatic_relation}), holds for the trapped system, the relation between the loss rate and the contact differs from that for the homogeneous system since the loss-rate coefficient of the trapped system is not only governed by particles being lost from the trap but also by a so-called deformation effect~\cite{gao2023temperaturedependent}.

\section{Validity Regime of the Theory based on $b_2$}
\label{SecIX}
\begin{figure*}[t!]
    \centering
    \includegraphics[width=0.8\textwidth]{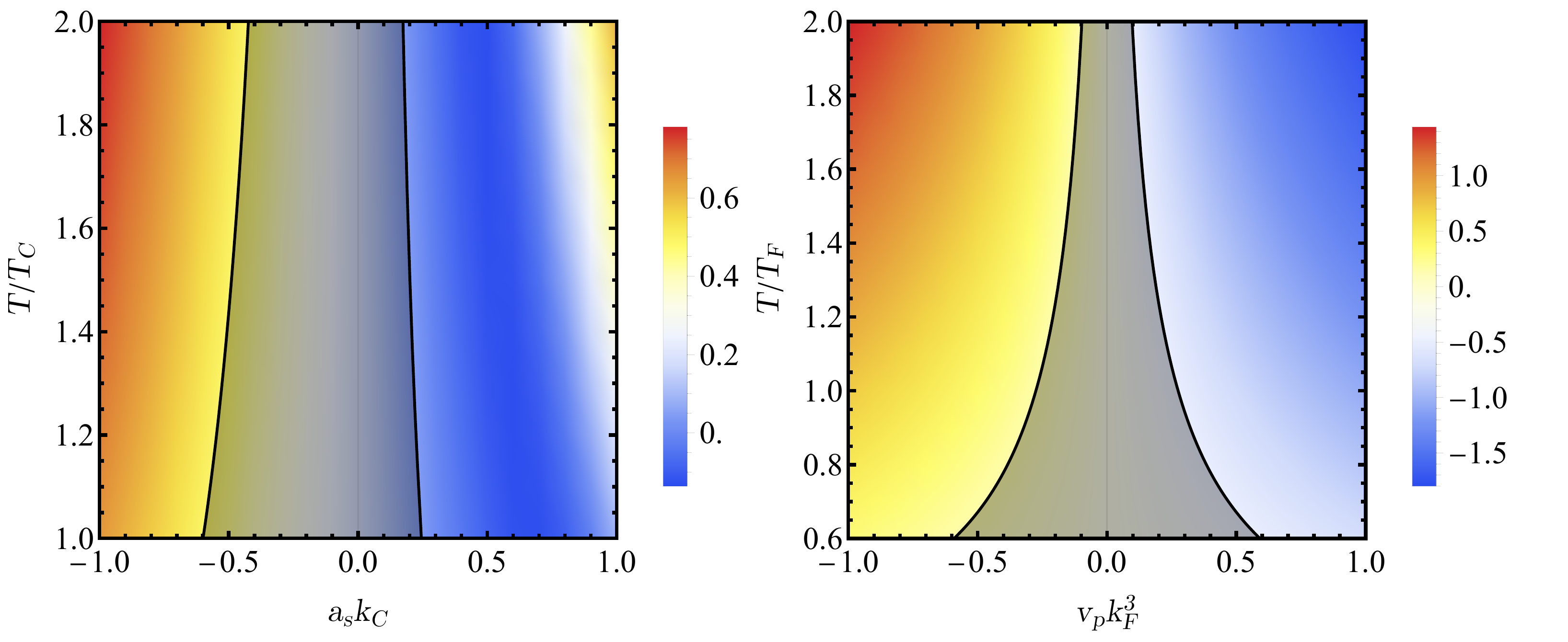}
    \caption{Validity regime determined by comparing the mean-field $b_2^{\text{mf}}$ and the exact $b_2$. Left panel: $s$-wave Bose gas. Right panel: $p$-wave Fermi gas. The heatmaps show the value of the exact $b_2$ [Eqs.~(\ref{s-wave_b2}) and (\ref{p-wave_b2})]; the color scheme is defined in the bars that are shown to the right of the main panels. The gray-shaded areas denote the parameter regime for which $|b_2-b_2^{\text{mf}}|/|b_2| \le 0.05$. A normalized difference below 5~\% is interpreted as an indicator that the theory framework developed in this work is applicable.}
    \label{SMfig1}
\end{figure*}

The results presented in this paper employ the mean-field framework and 
expansions applicable to the weakly interacting regime.
It is thus natural to ask what the validity regime of the theory is and whether the theory covers the operating regime of typical state-of-the-art experiments. 

The applicability regime can be estimated using the virial EOS of the homogeneous system that accounts for the second-order virial coefficient. At this level, the virial EOS can be analytically tackled for an arbitrary interaction strength. The virial EOS up to $b_2$ allows us to obtain exact reference results that can be used to assess the accuracy of Eq.~(\ref{Omega}). To facilitate the comparison, we compare the second-order virial coefficient $b_2$, obtained within the mean-field framework, directly with its exact counterpart.

We consider the $s$-wave Bose and $p$-wave Fermi gases as examples. Their exact $b_2$ are
\begin{equation}
    b_2^{\text{bose}}=\begin{cases}
    \frac{1}{4\sqrt{2}}+2\sqrt{2}\exp\left(-\frac{2}{\Tilde{a}_s^2\Tilde{T}}\right)\\
    -\sqrt{2}\exp\left(\frac{2}{\Tilde{a}_s^2\Tilde{T}}\right)\mathrm{erfc}\left(\sqrt{\frac{2}{\Tilde{T}}}\frac{1}{\Tilde{a}_s}\right)~~~\Tilde{a}_s>0\\
    \frac{1}{4\sqrt{2}}+\sqrt{2}\exp\left(\frac{2}{\Tilde{a}_s^2\Tilde{T}}\right)\\
    +\sqrt{2}\exp\left(\frac{2}{\Tilde{a}_s^2\Tilde{T}}\right)\mathrm{erf}\left(\sqrt{\frac{2}{\Tilde{T}}}\frac{1}{\Tilde{a}_s}\right)~~~\Tilde{a}_s<0\\
    \end{cases},
    \label{s-wave_b2}
\end{equation}
where $\Tilde{a}_s$ is equal to $a_s k_C$ and $\Tilde{T}$ is equal to $T/T_C$, and
\begin{equation}
\begin{split}
b_2^{\text{fermi}}=&-\frac{1}{4\sqrt{2}}+3\sqrt{2}\exp\left(\frac{2}{\Tilde{T}|\Tilde{v}_p|^{2/3}}\right)\\
&-6\sqrt{2}\exp\left(-\frac{1}{\Tilde{T}|\Tilde{v}_p|^{2/3}}\right)\cos\left(\frac{\sqrt{3}}{\Tilde{T}|\Tilde{v}_p|^{2/3}}\right)\\
&-\frac{48{}_1F_3\left[1;\frac{5}{6},\frac{7}{6},\frac{3}{2};\frac{8}{27\Tilde{T}^3\Tilde{v}_p^2}\right]}{\sqrt{\pi}\Tilde{T}^{3/2}\Tilde{v}_p},
\end{split}
\label{p-wave_b2}
\end{equation}
where $\Tilde{v}_p$ is equal to $v_p k_F^3$ and $\Tilde{T}$ is equal to $T/T_F$. $\mathrm{erf}$, $\mathrm{erfc}$, and ${}_pF_q$ denote the error, complementary error, and generalized hypergeometric functions, respectively. Equations~(\ref{s-wave_b2}) and (\ref{p-wave_b2}) are obtained from the Beth-Uhlenbeck formula~\cite{liu2013virial}. The expressions are expected to apply to relatively large $|\Tilde{a}_s|$ or $|\Tilde{v}_p|$. In writing Eq.~(\ref{p-wave_b2}), we ignored the contribution from the shallow bound state, i.e., we assumed that the effective range is infinitely large. Using Eq.~(\ref{virial_coefficients}), the second-order mean-field level virial coefficients, extracted from the EOS for $l=0$ and $l=1$ derived in this work, read
\begin{align}
    b_2^{\text{bose,mf}}&=\frac{1}{4\sqrt{2}}-\frac{\Tilde{a}_s\Tilde{T}}{\sqrt{\pi}},\label{s-wave_b2_mf}\\
    b_2^{\text{fermi,mf}}&=-\frac{1}{4\sqrt{2}}-\frac{9\Tilde{T}^{3/2}\Tilde{v}_p}{4\sqrt{\pi}}.\label{p-wave_b2_mf}
\end{align}

Figure~\ref{SMfig1} shows the virial coefficients $b_2^{\text{bose}}$ (left panel) 
and $b_2^{\text{fermi}}$
(right panel) as functions of the reduced interaction strength and reduced temperature. If the normalized difference $|b_2-b_2^\text{mf}|/|b_2|$between the mean-field coefficient and the exact virial coefficient is small, the mean-field treatment is expected to provide a faithful description. The gray-shaded area in Fig.~\ref{SMfig1}  demarcates the parameter combinations for which the normalized difference $|b_2-b_2^\text{mf}|/|b_2|$ between the mean-field virial coefficient and the exact virial coefficient is smaller than $0.05$. A difference below 5~\% is interpreted as indicating that the mean-field description provides an accurate description of the system. 
 At  $\Tilde{T}\simeq2$,
which is typical for experiments, the normalized difference is smaller than 5~\%  for  $|\Tilde{a}_s|\lesssim 0.15$ and $|\Tilde{v}_p|\lesssim0.1$ for the $s$-wave Bose gas and $p$-wave Fermi gas, respectively. For lower values of $\Tilde{T}$, the validity regime of the mean-field treatment extends, according to our criteria, to larger reduced interaction strengths. We caution that---even though the validity regime, as determined by $|b_2-b_2^\text{mf}|/|b_2|$---is quite large for temperatures that are much lower than the degeneracy temperature, Fig.~\ref{SMfig1} may not accurately reflect the validity regime of the full EOS. The reason is that the virial expansion fails at these low temperatures for which the fugacity is large (in this regime, virial coefficients other than $b_2$ come into play). We expect the ``true" validity range for the mean-field EOS to be comparable to that at temperatures notably above the transition temperature.

\begin{table*}[t]
    \centering
    \setlength{\tabcolsep}{10pt} 
\renewcommand{\arraystretch}{1.5} 
    \begin{tabular}{|c|c|c|c|}
\hline
molecule & statistics & interaction strength & reduced interaction strength \\ \hline
${}^{87}\mathrm{Rb}{}^{133}\mathrm{Cs}$~\cite{gregory2019sticky}&Boson&$a_s\approx 233a_0$~\cite{gregory2019sticky} & $\Tilde{a}_s\approx 0.02$\\ \hline
${}^{40}\mathrm{K}{}^{87}\mathrm{Rb}$~\cite{demarco2019degenerate}&Fermion&$v_p\approx(118a_0)^3$~\cite{idziaszek2010universal} & $\Tilde{v}_p\approx 1.44\times10^{-5}$\\ \hline
${}^{23}\mathrm{Na}{}^{40}\mathrm{K}$~\cite{duda2023longlived}&Fermion&$v_p\approx(88a_0)^3$~\cite{duda2023longlived} & $\Tilde{v}_p\approx 2.39\times10^{-7}$\\ \hline
\end{tabular}
\caption{Species used in molecular gas experiments, their statistics, their interaction strength, and their reduced interaction strength.}
\label{tab2}
\end{table*}

Ultracold molecular gas experiments typically work with samples that are characterized by extremely weak interactions, which are thus expected to be well described by the theory developed in this work. Table~\ref{tab2} shows three examples: one for a $s$-wave Bose gas and two for $p$-wave Fermi gases. The molecular gas is loaded into an approximately harmonic trap in the experiments. We use the typical densities reported in the experimental  works~\cite{gregory2019sticky,demarco2019degenerate,duda2023longlived} to calculate the reduced interaction strength and subsequently calculate the transition temperature $T_C$ and Fermi temperature $T_F$ assuming that the systems are homogeneous.
Table~\ref{tab2} shows that the reduced interaction strength is well within the parameter regime where the mean-field-based theory developed in this work is applicable.

\section{Conclusion}
\label{SecX}

In summary, we theoretically derived the EOS for single-component normal-phase quantum gases, which can be used to obtain all thermodynamic quantities. We focused on the behavior of the contact of two commonly produced systems---$s$-wave Bose and $p$-wave Fermi gases. We showed that the contact is purely a two-body quantity in the former system and exhibits many-body characteristics in the latter. We analyzed the behavior of the $p$-wave contact in the near-degenerate regime and found that the three-body contribution plays a vital role. The discussion was extended to harmonically trapped systems, where we analyzed the contacts under the LDA.

Our study provides critically needed guidance for recent ultracold molecular gas experiments, i.e., for weakly interacting molecules in the deeply degenerate regime where the virial expansion fails. Specifically, our results can be used to calibrate loss rate and temperature measurements. Moreover, our results also apply to single-component Fermi gases such as $^6$Li and $^{40}$K~\cite{venu2023unitarya,gerken2019observation}, and provide a reference for studying crossover from weakly to strongly interacting systems.\break

\section{Acknowledgement}

This work is supported by the National Natural Science Foundation of China under Grant No. 12204395, Hong Kong RGC Early Career Scheme (Grant No. 24308323) and Collaborative Research Fund (Grant No. C4050-23GF) the Space Application System of China Manned Space Program, and CUHK Direct Grant No. 4053676. D.B. acknowledges support from the National Science Foundation (NSF) through Grant No. PHY-2110158.\break

\appendix

\section{Bare Coupling and Scattering Length}
\label{appendix:A}

To obtain the relation between the bare couplings $g_l$ and the scattering lengths $a_l$ for each partial wave channel $l$, we follow the standard renormalization procedure, i.e., we  compare the $T$-matrix element $T_l(\mathbf{k},\mathbf{k}')$ and the partial wave scattering amplitude $f_l$~\cite{pethick2008bose}:
\begin{equation}
    -\frac{M V}{4\pi\hbar^2}T_l(\mathbf{k},\mathbf{k}')=(2l+1)f_l(k)P_l(\cos{\theta}),
    \label{SM::Tmatrix_comparison}
\end{equation}
where
\begin{equation}
    f_l(k)=\frac{1}{k\cot{\delta_l(k)}-ik}.
\end{equation}
In Eq.~(\ref{SM::Tmatrix_comparison}), $M$, $V$, and $k$ denote the mass of the constituent, volume, and relative wave vector, respectively. $P_l$ is the Legendre polynomial of degree $l$. The $T$-matrix elements are defined by the Schwinger-Dyson equation~\cite{sakurai1994modern}
\begin{equation}
\begin{split}
    &T_l(\mathbf{k}_1,\mathbf{k}_2)=U_l(\mathbf{k}_1,\mathbf{k}_2)\\
    &+V \int d^3 q {U}_l(\mathbf{k}_1,\mathbf{q})\frac{1}{\frac{\hbar^2 k_2^2}{M}-\frac{\hbar^2 q^2}{M}+i\epsilon}T_l(\mathbf{q},\mathbf{k}_2),
    \label{SM::schwinger_dyson_eqn}
\end{split}
\end{equation}
where $\epsilon$ has an infinitesimally small positive real value that ensures retarded propagation. Making use of the form of interaction in the main text, Eq.~(\ref{SM::schwinger_dyson_eqn}) can be worked out explicitly:
\begin{equation}
\begin{split}
\label{eq_help2}
    &T_l(\mathbf{k}_1,\mathbf{k}_2)=\frac{4\pi g_l}{V}\sum_{m=-l}^lk_1^lk_2^l{Y}_{lm}(\hat{\mathbf{k}}_1){Y}^*_{lm}(\hat{\mathbf{k}}_2)\\
    &\times\left[1+\frac{g_l}{2\pi^2}\int d q \frac{q^{2l+2}}{\frac{\hbar^2k^2}{M}-\frac{\hbar^2q^2}{M}+i\epsilon}+\cdots\right].
\end{split}
\end{equation}
We find that the infinite sum inside the square brackets forms a geometric sequence of the form $1+c+c^2+\cdots$, where $c$ is equal to the second term in square brackets in the last line of Eq.~(\ref{eq_help2}).
Hence, we can write
\begin{equation}
    T_l(\mathbf{k}_1,\mathbf{k}_2)=\dfrac{\frac{4\pi}{V}\sum_{m=-l}^lk_1^lk_2^l{Y}_{lm}(\hat{\mathbf{k}}_1){Y}^*_{lm}(\hat{\mathbf{k}}_2)}{\frac{1}{g_l}-\frac{1}{2\pi^2}\int_0^\infty d q\frac{q^{2l+2}}{\frac{\hbar^2k^2}{M}-\frac{\hbar^2q^2}{M}+i\epsilon}}.
\end{equation}
Because of energy conservation, the magnitudes of $\mathbf{k}_1$ and $\mathbf{k}_2$ should be the same. Setting $|\mathbf{k}_1|=|\mathbf{k}_2|=k$, denoting the angle between $\mathbf{k}_1$ and $\mathbf{k}_2$ by $\theta$, and using the addition theorem  
\begin{equation}
    P_l(\hat{\mathbf{k}}_1 \cdot \hat{\mathbf{k}}_2)=\frac{4\pi}{(2l+1)}\sum_{m=-l}^l{Y}_{lm}(\hat{\mathbf{k}}_1){Y}^*_{lm}(\hat{\mathbf{k}}_2)
\end{equation}
of spherical harmonics,
we obtain
\begin{equation}
    T_l(\mathbf{k}_1,\mathbf{k}_2)=\dfrac{(2l+1)P_l(\cos\theta)}{\frac{V}{k^{2l}g_l}-\frac{V}{2k^{2l}\pi^2}\int_0^\infty dq\frac{q^{2l+2}}{\frac{\hbar^2k^2}{M}-\frac{\hbar^2q^2}{M}+i\epsilon}}.
\end{equation}
Combining Eq.~(\ref{SM::Tmatrix_comparison}) and the definition of the partial-wave phase shifts $\delta_l(k)$, we get
\begin{equation}
\label{eq_help1}
    \frac{1}{g_l}-\frac{1}{2\pi^2}\int_0^\infty dq \frac{q^{2l+2}}{\frac{\hbar^2k^2}{M}-\frac{\hbar^2q^2}{M}+i\epsilon}=\frac{M}{4\pi\hbar^2a_l}+\frac{iMk^{2l+1}}{4\pi\hbar^2}.
\end{equation}
The integral on the left-hand side of Eq.~(\ref{eq_help1}) diverges. It can be reexpressed using the well-known low-energy relation~\cite{sakurai1994modern}:
\begin{equation}
\begin{split}
\label{eq_help3}
    &\frac{1}{2\pi^2}\int_0^\infty dq \frac{q^{2l+2}}{\frac{\hbar^2k^2}{M}-\frac{\hbar^2q^2}{M}+i\epsilon}\\
    &=\frac{M}{2\pi^2\hbar^2}\int_{0}^{\infty}dq\mathcal{P}\frac{q^{2l+2}}{k^2-q^2}\\
    &-\frac{i\pi M}{2\pi^2\hbar^2}\int_{0}^{\infty}dqq^{2l+2}\delta(k^2-q^2)\\
    &\xrightarrow[]{k\rightarrow0}-\frac{M}{2\pi^2\hbar^2}\int_{0}^{\infty}dqq^{2l}+\frac{iMk^{2l+1}}{4\pi\hbar^2},
\end{split}
\end{equation}
where $\mathcal{P}$ denotes the Cauchy principal value. Using Eq.~(\ref{eq_help3}) in Eq.~(\ref{eq_help1}), we finally obtain the renormalization condition Eq.~(\ref{renormalization_condition}).
It is known that the derived renormalization condition cannot eliminate all diverging terms that may arise in the many-body treatment of the single-species bosonic system, especially when the interaction is strong~\cite{zhai2021ultracold}. 
For example, in single-component bosonic systems, Efimov physics requires one to introduce an additional parameter for renormalization~\cite{braaten2007efimov}. However,  since the lowest-order mean-field interactions dominate the many-body physics for the weakly-interacting systems of interest in this work, the renormalization condition is not needed to eliminate divergencies, i.e., we can use the non-integral part of the bare coupling constant.

\section{Details of Calculation on Thermodynamics}
\label{appendix:B}

\subsection{Self-Energy and Grand Potential}
\label{appendix:B1}

Substituting Eq.~(\ref{interaction}) into Eq.~(\ref{HF_approx}), the self-energy is found explicitly to be
\begin{equation}
 \label{eq_help6}
\begin{split}
     \hbar \Sigma_l(\mathbf{k})=&\frac{2^{1-2l}(2l+1)}{\pi}\frac{\hbar^2a_l}{M}\int_0^\infty dk' n_{\mathbf{k}'}(k')^2\\
     &\times\int_0^\pi d\theta|\mathbf{k}-\mathbf{k}'|^{2l}\sin(\theta),
\end{split}
\end{equation}
where $\theta$ denotes the angle between $\mathbf{k}$ and $\mathbf{k}'$.
To proceed, we work in the weakly-interacting limit and assume that $n_\mathbf{k}$ is equal to the non-interacting  distribution functions  $n^{(0)}_\mathbf{k}$,
\begin{equation}
    \label{eq_help5}
n^{(0)}_\mathbf{k}=\left[\exp\left(\dfrac{\epsilon_k^{(0)}}{k_B T}\right)z^{-1}\mp1\right]^{-1},
\end{equation}
Inserting Eq.~(\ref{eq_help5}) into Eq.~(\ref{eq_help6}), we note that the $\theta$ dependence only appears in the term $|\mathbf{k}-\mathbf{k}'|^{2l}$. 
To evaluate the integral over $\theta$, the usual binomial theorem for scalars cannot be applied to $|\mathbf{k}-\mathbf{k}'|^{2l}$. Instead, we first take the square and then construct a series expansion using  the trinomial theorem~\cite{weissteintrinomial}:
\begin{equation}
\begin{split}
    |\mathbf{k}-\mathbf{k}'|^{2l}&=[k^2+(k')^2-2\mathbf{k}\cdot\mathbf{k}']^l\\
    &=\sum_{i,j,n \atop i+j+n=l}\frac{l!}{i!j!n!}(-2)^nk^{2i+n}(k')^{2j+n}\cos^n\theta,
\end{split}
\end{equation}
where the indices $i,j,$ and $n$ each go from $0$ to $l$. Letting, as before, the angle between $\mathbf{k}$ and $\mathbf{k}'$ be $\theta$ and using
\begin{equation}
    \int_0^\pi\cos(\theta)^n\sin(\theta)d\theta=\frac{1+(-1)^n}{1+n},~~~n=0,1,2,3,\cdots,
\end{equation}
we have
\begin{equation}
\begin{split}
    &\hbar \Sigma_l(\mathbf{k})\xrightarrow[]{|a_lk_F^{2l+1}|\ll1}\frac{2^{1-2l}(2l+1)}{\pi}\frac{\hbar^2a_l}{M}\\
    &\times\sum_{i,j,n \atop i+j+n=l}\frac{l!}{i!j!n!}2^n\frac{1+(-1)^n}{1+n}k^{2i+n}\int dk' n^{(0)}_{\mathbf{k}'} (k')^{2j+n+2}.
    \label{SM::self_energy}
\end{split}
\end{equation}
Using the integral expression of the polylogarithm function, 
\begin{equation}
    \int dq n_\mathbf{q}^{(0)} q^j=\pm  2^{\frac{j-1}{2}}\left(\frac{M k_B T}{\hbar^2}\right)^{\frac{j+1}{2}}\Gamma\left(\frac{j+1}{2}\right)\mathrm{Li}_{\frac{j+1}{2}}(\pm  z),
    \label{SM::polylog_int}
\end{equation}
the self-energy becomes
\begin{equation}
\begin{split}
    &\hbar \Sigma_l(\mathbf{k})\approx\pm\frac{2l+1}{\pi}\frac{\hbar^2a_l}{M}2^{\frac{3}{2}-2l}\\
    &\times\sum_{i,j,n \atop i+j+n=l}\frac{l!}{i!j!n!}\frac{1+(-1)^n}{1+n}k^{2i+n}2^{j+\frac{3n}{2}}\\
    &\Gamma\left(\frac{2j+n+3}{2}\right)\left(\frac{M k_B T}{\hbar^2}\right)^{\frac{2j+n+3}{2}}\mathrm{Li}_{\frac{2j+n+3}{2}}(\pm z).
    \label{SM::self-energy}
\end{split}
\end{equation}
Equation~(\ref{SM::self-energy}) reveals the structure of the self-energy for the $l$th partial-wave channel clearly: $\Sigma_l(\mathbf{k})=\Sigma_l(k)$ is a polynomial of even degree in $k$ (see the factor of  $k^{2i+n}$ in the summand) since $2i$ is always even and the summand is zero when $n$ is odd.
For example, $\Sigma_0=A_0$ for the $s$-wave channel, $\Sigma_1=A_1+B_1 k^2$ for the $p$-wave channel,  $\Sigma_2=A_2+B_2 k^2+C_2 k^4$ for the $d$-wave channel, etc., where $A_l$, $B_l$, and $C_l$ are constants that depend on the temperature $T$.

Since $\Sigma_l(\mathbf{k},\lambda)$ is directly proportional to $\lambda$ and $a_l$ [see Eq.~(\ref{SM::self-energy})] and since $n_{\mathbf{k}}(\lambda)$ has no dependence on $\lambda$ at leading order [$n_{\mathbf{k}}(\lambda)\approx n_\mathbf{k}^{(0)}$], the leading-order modification of the grand potential based on Eq.~(\ref{Hellman-Feynman_theorem}) is given by
\begin{equation}
    \Omega-\Omega^{(0)}\approx \frac{V}{4\pi^2} \int dk k^2\hbar \Sigma_l(\mathbf{k})n^{(0)}_\mathbf{k}.
\end{equation}
Using Eqs.~(\ref{SM::self_energy}) and (\ref{SM::polylog_int}), we arrive at Eq.~(\ref{Omega}) in the main text.

For convenience, we list the explicit expressions of Eq.~(\ref{Omega}) from the main text for $l=0$ to $l=3$:
\begin{widetext}
\begin{align}
    \frac{\Omega_0}{k_B T V}&=-\frac{\mathrm{Li}_{5/2}(z)}{\lambda_T^3}+\frac{2a_0[\mathrm{Li}_{3/2}(z)]^2}{\lambda_T^4}~(s\mathrm{-wave, Bose~gas}),\\
    \frac{\Omega_1}{k_B T V}&=\frac{\mathrm{Li}_{5/2}(-z)}{\lambda_T^3}+\frac{18\pi a_1\mathrm{Li}_{3/2}(-z)\mathrm{Li}_{5/2}(-z)}{\lambda_T^6}~(p\mathrm{-wave, Fermi~gas}),\\
    \frac{\Omega_2}{k_B T V}&=-\frac{\mathrm{Li}_{5/2}(z)}{\lambda_T^3}+\frac{75\pi^2a_2[\mathrm{Li}_{5/2}(z)]^2}{\lambda_T^8}+\frac{75\pi^2a_2\mathrm{Li}_{3/2}(z)\mathrm{Li}_{7/2}(z)}{\lambda_T^8}~(d\mathrm{-wave, Bose~gas}),\\
    \frac{\Omega_3}{k_B T V}&=\frac{\mathrm{Li}_{5/2}(-z)}{\lambda_T^3}+\frac{2205\pi^3a_3\mathrm{Li}_{5/2}(-z)\mathrm{Li}_{7/2}(-z)}{2\lambda_T^{10}}+\frac{735\pi^3a_3\mathrm{Li}_{3/2}(-z)\mathrm{Li}_{9/2}(-z)}{2\lambda_T^{10}}~(f\mathrm{-wave, Fermi~gas}).
\end{align}
\end{widetext}

\subsection{Isothermal Compressibility}

 This section considers the isothermal compressibility, which is defined through
\begin{equation}
    \kappa_T=\frac{z}{k_B T}\frac{\partial n}{\partial z}.
\end{equation}
In terms of the grand potential $\Omega$, we find
\begin{equation}
    \kappa_T=-\frac{z}{N k_B T}\frac{\partial(\Omega/k_B T)}{\partial z}-\frac{z^2}{Nk_B T}\frac{\partial^2(\Omega/k_B T)}{\partial z^2}.
\end{equation}
Using Eq.~(\ref{Omega}), we find the isothermal compressibility in terms of the fugacity:
\begin{equation}
\begin{split}
    &\kappa_T=\pm \frac{\mathrm{Li}_{1/2}(\pm z)}{n k_B T \lambda_T^3}-\frac{2}{n k_B T\lambda_T^3}\frac{a_l}{\lambda_T^{2l+1}}\sum_{i,j,n\atop i+j+n=l}^l\mathcal{C}(i,j,n,l)\\
    &\times\left[\mathrm{Li}_{\frac{2i+n+1}{2}}(\pm z)\mathrm{Li}_{\frac{2j+n+1}{2}}(\pm z)+\mathrm{Li}_{\frac{2i+n-1}{2}}(\pm z)\mathrm{Li}_{\frac{2j+n+3}{2}}(\pm z)\right].
\end{split}
\label{kappa_z}
\end{equation}
This expression is not yet in the ``standard form"  of the compressibility, as it depends on the fugacity (a thermodynamic variable in the grand canonical potential) rather than the density
(a thermodynamic variable in the canonical ensemble). To convert $z$ to $n$, we use the Gibbs-Duhem relation
\begin{equation}
\begin{split}
    n=-z\frac{\partial(\Omega/k_B T V)}{\partial z}.
\end{split}
\end{equation}
We know that $n$ is equal to $\pm \mathrm{Li}_{3/2}(\pm z^{(0)})/\lambda_T^3$ at leading order. If we write $z$ in terms of $z^{(0)}$,  $z=z^{(0)}+\delta z$, then we find to first order,
\begin{equation}
\begin{split}
    &\delta z(|a_l|/\lambda_T^{2l+1}\rightarrow0)=\pm z_0\frac{2a_l}{\lambda_T^{2l+1}}\sum_{i,j,n\atop i+j+n=l}\mathcal{C}(i,j,n,l)\\
    &\times\mathrm{Li}_{\frac{2i+n+1}{2}}(\pm z^{(0)})\mathrm{Li}_{\frac{2j+n+3}{2}}(\pm z^{(0)})/{\mathrm{Li}_{1/2}(\pm z^{(0)})}.
\end{split}
\end{equation}
Substituting $z=z^{(0)}+\delta z$ into Eq.~(\ref{kappa_z}), we have
\begin{equation}
\begin{split}
    n k_B T\kappa_T=&\pm \frac{\mathrm{Li}_{1/2}(\pm z^{(0)})}{\lambda_T^3}-\frac{2a_l}{\lambda_T^{2l+4}}\sum_{i,j,n\atop i+j+n=l}\mathcal{C}(i,j,n,l)\\
    &\times\big(\mathrm{Li}_{\frac{2i+n+1}{2}}(\pm z^{(0)})\mathrm{Li}_{\frac{2j+n+1}{2}}(\pm z^{(0)})\\
    &+\mathrm{Li}_{\frac{2i+n-1}{2}}(\pm z^{(0)})\mathrm{Li}_{\frac{2j+n+3}{2}}(\pm z^{(0)})\\
    &-\mathrm{Li}_{-\frac{1}{2}(\pm z^{(0)})}{\mathrm{Li}_{\frac{2i+n+1}{2}}}(\pm z^{(0)})\\
    &\times{\mathrm{Li}_{\frac{2j+n+3}{2}}(\pm z^{(0)})}/{\mathrm{Li}_{\frac{1}{2}}(\pm z^{(0)})}\big).
\end{split}
\end{equation}
The explicit expressions for $l=0$ and $l=1$ read
\begin{equation}
\begin{split}
    n k_B T\kappa_T&=\frac{\mathrm{Li}_{1/2}(z^{(0)})}{\lambda_T^3}-\frac{4a_0[\mathrm{Li}_{1/2}(z^{(0)})]^2}{\lambda_T^4}~~~\mathrm{for}~l=0,\label{SM::swave_bose_compressibility}\\
    n k_B T\kappa_T&=-\frac{\mathrm{Li}_{1/2}(-z^{(0)})}{\lambda_T^3}+\frac{54\pi a_1 n\mathrm{Li}_{1/2}(-z^{(0)})}{\lambda_T^3}\\
    &+\frac{18\pi a_1 n^2\mathrm{Li}_{-1/2}(-z^{(0)})}{\mathrm{Li}_{1/2}(-z^{(0)})}~~~\mathrm{for}~l=1.
\end{split}
\end{equation}
The divergence of the compressibility of the bosonic $s$-wave system indicates the Bose-Einstein condensate (BEC) transition. The first equation of Eqs.~(\ref{SM::swave_bose_compressibility}) diverges at $z^{(0)}=1$ or $T=T_C$, i.e., the BEC transition temperature of the non-interacting system. The argument extends to Bose gases with higher partial-wave interactions because $\mathrm{Li}_{s}(x)$ diverges at $x=1$ for all $s$, $s\leq1$.  It is worthwhile pointing out that $T_C$ does, in fact, have a correction of order $n^{1/3}a_0$, which arises from higher-order fluctuations that are not captured by the mean-field approach considered here~\cite{andersen2004theory}.

\subsection{Entropy and Isochoric Heat Capacity}

This section determines the isochoric heat capacity. We start with the entropy, which is defined through
\begin{equation}
\begin{split}
    S&=-\left.\frac{\partial \Omega}{\partial T}\right|_\mu=-\left(\left.\frac{\partial \Omega}{\partial T}\right|_z+\left.\frac{\partial \Omega}{\partial z}\right|_T \left.\frac{\partial z}{\partial T}\right|_\mu \right)\\
    &=-\left.\frac{\partial \Omega}{\partial T}\right|_z+\frac{z\ln(z)}{T}\left.\frac{\partial \Omega}{\partial z}\right|_T.
\end{split}
\end{equation}
To obtain the entropy in the canonical ensemble in terms of 
 $n$ and $T$, we change $z$ to $z^{(0)}$ and keep terms up to  first order in $a_l/\lambda_T^{2l+1}$:
\begin{equation}
\begin{split}
    S&=\frac{k_BV}{2\lambda_T^3}\left[\pm5\mathrm{Li}_{5/2}(\pm z^{(0)})\mp2\ln(z^{(0)})\mathrm{Li}_{3/2}(\pm z^{(0)})\right]\\
    &+\frac{k_BVa_l}{\lambda_T^{2l+4}}\sum_{i,j,n\atop i+j+n=l}\mathcal{C}(i,j,l,n)\big[3{\mathrm{Li}_{3/2}(\pm z^{(0)})}\\
    &\times\mathrm{Li}_{\frac{2i+n+1}{2}}(\pm z^{(0)})\mathrm{Li}_{\frac{2j+n+3}{2}}(\pm z^{(0)})/{\mathrm{Li}_{\frac{1}{2}}(\pm z^{(0)})}\\
    &-(l+3)\mathrm{Li}_{\frac{2i+n+3}{2}}(\pm z^{(0)})\mathrm{Li}_{\frac{2j+n+3}{2}}(\pm z^{(0)})\big].
\end{split}
\end{equation}
It can be noted that the entropy for $l=0$ is quite special since it is independent of $a_0$. After simplification, we find that the entropy of the weakly-interacting $s$-wave Bose gas is identical to that of the non-interacting Bose gas: 
\begin{equation}
    S_{l=0}=\frac{k_BV}{2\lambda_T^3}\left[5\mathrm{Li}_{5/2}(z^{(0)})-2\ln(z^{(0)})\mathrm{Li}_{3/2}(z^{(0)})\right].
\end{equation}
 The independence of $a_0$ is a consequence of the fact that the $s$-wave mean-field interaction simply shifts the chemical potential, making the interacting system resemble a non-interacting gas. The $p$-wave interactions, in contrast, modify the entropy of the non-interacting system: 
\begin{equation}
\begin{split}
    S_{l=1}=&\frac{k_BV}{2\lambda_T^3}\left[-5\mathrm{Li}_{5/2}(-z^{(0)})+2\ln(z^{(0)})\mathrm{Li}_{3/2}(-z^{(0)})\right]\\
    &+\frac{k_BV}{\lambda_T^3}\frac{27\pi a_1 [\mathrm{Li}_{\frac{3}{2}}(-z^{(0)})]^3}{\lambda_T^3\mathrm{Li}_{\frac{1}{2}}(-z^{(0)})}\\
    &-\frac{k_BV}{\lambda_T^3}\frac{45\pi a_1\mathrm{Li}_{\frac{3}{2}}(-z^{(0)})\mathrm{Li}_{\frac{5}{2}}(-z^{(0)})}{\lambda_T^3}.
\end{split}
\end{equation}
We can now calculate the isochoric heat capacity directly from the entropy,
\begin{equation}
\begin{split}
    C_V=&T\left.\frac{\partial S}{\partial T}\right|_{n}=T\frac{\partial \lambda_T}{\partial T}\left(\left.\frac{\partial S}{\partial \lambda_T}\right|_{z^{(0)}}+\left.\frac{\partial S}{\partial z^{(0)}}\right|_{\lambda_T}\left.\frac{\partial z^{(0)}}{\partial \lambda_T}\right|_{n}\right)\\
    &=-\frac{\lambda_T}{2}\left(\left.\frac{\partial S}{\partial \lambda_T}\right|_{z^{(0)}}\pm\frac{3n\lambda_T^2z^{(0)}}{\mathrm{Li}_{\frac{1}{2}}(\pm z^{(0)})}\left.\frac{\partial S}{\partial z^{(0)}}\right|_{\lambda_T}\right)
\end{split}
\end{equation}
Explicitly, the expression is
\begin{equation}
\begin{split}
    C_V=&\frac{k_B V}{\lambda_T^3}\left[\pm\frac{15\mathrm{Li}_{\frac{5}{2}}(\pm z^{(0)})}{4}\mp\frac{9 n^2 \lambda_T^6}{4\mathrm{Li}_{\frac{1}{2}}(\pm z^{(0)})}\right]\\
    &-\frac{k_B V a_l}{2\lambda_T^{2l+4}}\sum_{i,j,n\atop i+j+n=l}\frac{\mathcal{C}(i,j,n,l)}{[\mathrm{Li}_{\frac{1}{2}}(\pm z^{(0)})]^3}\\
    &\times\bigg\{2(l^2+5l+6)[\mathrm{Li}_{\frac{1}{2}}(\pm z^{(0)})]^3\\
    &\mathrm{Li}_{\frac{2i+n+3}{2}}(\pm z^{(0)})\mathrm{Li}_{\frac{2j+n+3}{2}}(\pm z^{(0)})+9n^2\lambda_T^6\mathrm{Li}_{\frac{1}{2}}(\pm z^{(0)})\\
    &\times\big[\mathrm{Li}_{\frac{2i+n+1}{2}}(\pm z^{(0)})\mathrm{Li}_{\frac{2j+n+1}{2}}(\pm z^{(0)})\\
    &+\mathrm{Li}_{\frac{2i+n-1}{2}}(\pm z^{(0)})\mathrm{Li}_{\frac{2j+n+3}{2}}(\pm z^{(0)})\big]\\
    &-(9n^2\lambda_T^6\mathrm{Li}_{-\frac{1}{2}}(\pm z^{(0)})\\
    &\mp3(4l+7)n\lambda_T^3[\mathrm{Li}_{\frac{1}{2}}(\pm z^{(0)})]^2)\\
    &\times\mathrm{Li}_{\frac{2i+n+1}{2}}(\pm z^{(0)})\mathrm{Li}_{\frac{2j+n+3}{2}}(\pm z^{(0)})\bigg\}.
\end{split}
\end{equation}
Consistent with the discussion above, one can check that the weakly interacting $s$-wave system has the same $C_V$ as the corresponding non-interacting system. In contrast, the isochoric heat capacity of the weakly-interacting $p$-wave gas contains a correction due to the interactions.

\section{Calculation of Contacts via Truncated Virial Series}
\label{appendix:C}

This work uses the virial expansion, truncated at order $j$, to analyze the $j$-body contribution to the contact and other thermodynamic quantities. A crucial point of the derivations is to consistently account for terms that contribute to the $j$-body physics. In particular, this implies that higher-order terms that go beyond $j$-body physics need to be excluded.

We start from the general virial expansion in the grand canonical ensemble, truncated at order $j$,  $\Omega=-k_B T Z_1\sum_{i=1}^{j}b_iz^i$. Here,
$Z_1$ denotes the one-body partition function for an as-of-yet unspecified system (could be the homogeneous or inhomogeneous system). 
Using the standard thermodynamics relation $N=-z\frac{\partial(\Omega/k_B T)}{\partial z}$, we find
\begin{equation}
    N=Z_1\sum_{i=1}^j ib_jz^i.\label{SM::truncated_N}
\end{equation}
From the expression for $N$ and the definition of the contact, it is straightforward to calculate the contact in the grand canonical ensemble at order $j$:
\begin{equation}
    C_l(z)=-\frac{2k_BTM[\mathrm{Re}(a_l)]^2Z_1}{(2l+1)\hbar^2}\sum_{i=1}^j\frac{\partial b_i}{\partial a_l}z^i.\label{SM::truncated_Clz}
\end{equation}
In what follows, we are interested in the contact in the canonical ensemble. To convert Eq.~(\ref{SM::truncated_Clz}) from the grand canonical to the canonical ensemble, we need to express $z$ in terms of $N$. This is achieved by applying the Lagrange inversion theorem ({see Sec.} 3.6.6 of Ref.~\cite{abramowitz1988handbook}) to Eq.~(\ref{SM::truncated_N}). To obtain the forth-order results presented in the main text, we need to include terms up to the fourth power of $N$,
\begin{equation}
\begin{split}
    z=&\frac{b_1 N}{Z_1}-\frac{2b_2N^2}{Z_1^2}+\frac{(8b_2^2-3b_3)N^3}{Z_1^3}\\
    &+\frac{(-40b_2^3+30b_2b_3-4b_4)N^4}{Z_1^4}.\label{SM::truncated_z}
\end{split}
\end{equation}
When substituting Eq.~(\ref{SM::truncated_z}) into Eq.~(\ref{SM::truncated_Clz}) and working at order $j$, one needs to truncate all terms at order $N^j$.
Even though the $z^2$ term, e.g., generates terms that scale as $N^2,\cdots, N^8$,
only the terms proportional to $N^2$, $N^3$, and $N^4$ are kept to obtain consistent forth-order results. The results reported below are obtained by additionally taking, consistent with the weak interaction regime assumption, the  $|a_l|/\lambda_T^{2l+1}\rightarrow 0$ limit.

The abovementioned strategy is critical for determining that the homogeneous system's $s$-wave contact has a pure two-body character. 
If we---incorrectly so---kept all terms in $N$ when going from the grand canonical to the canonical ensemble, we would obtain
\begin{equation}
    \begin{split}
    C_0=&\frac{8\pi[\mathrm{Re}(a_0)]^2}{V}N^2~~~(\text{exact result})\\
   C_0\stackrel{?}{=}&
   \frac{8\pi[\mathrm{Re}(a_0)]^2}{V}N^2-\frac{16\pi^{5/2}\hbar^3[\mathrm{Re}(a_0)]^2}{(M k_B T)^{3/2}V^2}N^3\\
   &+\left(\frac{40\pi^4\hbar^6[\mathrm{Re}(a_0)]^2}{(M k_B T)^{3}V^3}-\frac{128\pi^4\hbar^6[\mathrm{Re}(a_0)]^2}{3\sqrt{3}(M k_B T)^{3}V^3}\right)N^4\\
   &+\mathcal{O}(N^5)~~~(\text{truncated at $j=2$})\\
    C_0\stackrel{?}{=}&\frac{8\pi[\mathrm{Re}(a_0)]^2}{V}N^2\\
    &+\left(\frac{-8\pi^4\hbar^6[\mathrm{Re}(a_0)]^2}{(M k_B T)^{3}V^3}-\frac{128\pi^4\hbar^6[\mathrm{Re}(a_0)]^2}{3\sqrt{3}(M k_B T)^{3}V^3}\right)N^4\\
    &+\mathcal{O}(N^5)~~~(\text{truncated at $j=3$})\\
    C_0\stackrel{?}{=}&\frac{8\pi[\mathrm{Re}(a_0)]^2}{V}N^2+\mathcal{O}(N^5)~~~(\text{truncated at $j=4$}).
    \end{split}
\end{equation}
Converting the above expressions to expressions that use $k_C=(\frac{8\pi^{3/2}}{\zeta(3/2)}\frac{N}{V})^{1/3}$
and $T_C=\frac{\hbar^2k_C^2}{2Mk_B}$ [implying $N=\frac{M^{3/2}\zeta(3/2)V}{2\sqrt{2}\pi^{3/2}\hbar^3}T_C^{3/2}]$, it can be observed that this approach yields inconsistent results with regards to the order of $N$:
\begin{equation}
    \begin{split}
    \frac{C_0}{N[\mathrm{Re}(a_0)]^2k_C^3}&=\frac{\zeta(3/2)}{\sqrt{\pi}}~~~(\text{exact result})\\
    \frac{C_0}{N[\mathrm{Re}(a_0)]^2k_C^3}&=\frac{\zeta(3/2)}{\sqrt{\pi}}-\frac{\zeta(3/2)^2}{\sqrt{2\pi}}\sqrt{\frac{T_C^3}{T^3}}\\
    &+\frac{\zeta(3/2)^3}{8\sqrt{\pi}}\frac{T_C^3}{T^3}\\
    &+\mathcal{O}((T_C/T)^{9/2})~~~(\text{truncated at $j=2$})\\
    \frac{C_0}{N[\mathrm{Re}(a_0)]^2k_C^3}&=\frac{\zeta(3/2)}{\sqrt{\pi}}+\frac{\zeta(3/2)^3}{8\sqrt{\pi}}\frac{T_C^3}{T^3}\\
    &-\frac{2\zeta(3/2)^3}{3\sqrt{3\pi}}\frac{T_C^3}{T^3}\\
    &+\mathcal{O}((T_C/T)^{9/2})~~~(\text{truncated at $j=3$})\\
    \frac{C_0}{N[\mathrm{Re}(a_0)]^2k_C^3}&=\frac{\zeta(3/2)}{\sqrt{\pi}}+\mathcal{O}((T_C/T)^{9/2})\\
    &(\text{truncated at $j=4$}).
    \end{split}
\end{equation}
On the other hand, when following the correct approach that consistently keeps terms up to order $N^j$, we find: 
\begin{equation}
    \begin{split}
    C_0=&\frac{8\pi[\mathrm{Re}(a_0)]^2}{V}N^2~~~(\text{exact result})\\
    C_0=&\frac{8\pi[\mathrm{Re}(a_0)]^2}{V}N^2~~~(\text{truncated at $j=2$})\\
    C_0=&\frac{8\pi[\mathrm{Re}(a_0)]^2}{V}N^2~~~(\text{truncated at $j=3$})\\
    C_0=&\frac{8\pi[\mathrm{Re}(a_0)]^2}{V}N^2~~~(\text{truncated at $j=4$}).
    \end{split}
\end{equation}
Again, in units 
of $k_C$ and $T_C$, we find:
\begin{equation}
    \begin{split}
    &\frac{C_0}{N[\mathrm{Re}(a_0)]^2k_C^3}=\frac{\zeta(3/2)}{\sqrt{\pi}}~~~(\text{exact result})\\
    &\frac{C_0}{N[\mathrm{Re}(a_0)]^2k_C^3}=\frac{\zeta(3/2)}{\sqrt{\pi}}~~~(\text{truncated at $j=2$})\\
    &\frac{C_0}{N[\mathrm{Re}(a_0)]^2k_C^3}=\frac{\zeta(3/2)}{\sqrt{\pi}}~~~(\text{truncated at $j=3$})\\
    &\frac{C_0}{N[\mathrm{Re}(a_0)]^2k_C^3}=\frac{\zeta(3/2)}{\sqrt{\pi}}~~~(\text{truncated at $j=4$}).
    \end{split}
\end{equation}
The analysis above leads to the following important conclusion: When one includes terms up to $j=2$, the $s$-wave result is exact; contributions from larger clusters do not lead to any corrections.

\section{Zero-Temperature Limit in Fermionic Systems}
\label{appendix:D}

The main text characterizes weakly interacting single-component systems in the normal phase. This appendix provides limiting expressions for several quantities of fermionic systems in the $T/T_F \rightarrow 0$ limit.  In the $T/T_F \rightarrow 0$ limit, 
it is computationally inefficient to evaluate expressions that explicitly or implicitly contain polylogarithm functions numerically. 
The reason is that the direct use of $T=0$ yields expressions like $\text{Li}_s(-\infty)$ cannot be evaluated numerically.
Instead, it is generally more convenient to use asymptotic analytic expressions.  The large-argument asymptote of fermionic-type polylogarithm functions is~\cite{wood1992computation} 
\begin{equation}
    -\mathrm{Li}_s(-x)\xrightarrow[]{x\rightarrow+\infty}\frac{[\ln(x)]^s}{\Gamma(s+1)}.
    \label{SM::Li}
\end{equation}
Correspondingly, one finds
\begin{equation}
    -\mathrm{Li}_{s_1}[\mathrm{Li}^{-1}_{s_2}(-x)]\xrightarrow[]{x\rightarrow+\infty}\dfrac{[\Gamma(1+s_2)x]^{s_1/s_2}}{\Gamma(1+s_1)},
    \label{SM::LiLiinv}
\end{equation}
where $^{-1}$ denotes the inverse function.
Direct application of Eq.~(\ref{SM::LiLiinv}) to the homogeneous system yields
\begin{equation}
    \begin{split}
        \frac{C_1}{N[\mathrm{Re}(a_1)]^2k_F^5}&\xrightarrow[]{T\rightarrow0}\frac{2}{5\pi},\\
        \frac{C_3}{N[\mathrm{Re}(a_3)]^2k_F^9}&\xrightarrow[]{T\rightarrow0}\frac{4}{45\pi}.
    \end{split}
\end{equation}
One also needs to change the integration limits accordingly for harmonically trapped systems. To illustrate this, we consider the $p$-wave system. The exact integral for the contact is
\begin{equation}
\begin{split}
    \frac{C_1^\mathrm{trap}}{N[\mathrm{Re}(a_1)]^2(k_F^{\mathrm{trap}})^5}=&\int_0^\infty d\Bar{r} \frac{18\Bar{T}^4}{\pi} \Bar{r}^2\mathrm{Li}_{3/2}\left(-e^{-\frac{\Bar{r}^2}{\Bar{T}}}z^{(0)}\right)\\
    &\times\mathrm{Li}_{5/2}\left(-e^{-\frac{\Bar{r}^2}{\Bar{T}}}z^{(0)}\right),
\end{split}
\end{equation}
where $\Bar{T}=T/T_F^\mathrm{trap}$ and $\Bar{r}=r/R_F$. Here, $R_F=(48N)^{1/6}\sqrt{\hbar/M\omega}$ denotes the Thomas-Fermi radius. Utilizing Eq.~(\ref{SM::Li}) to simplify the integrand, we find
$
    \frac{18\Bar{T}^4}{\pi} \Bar{r}^2\mathrm{Li}_{3/2}\left(-e^{-\frac{\Bar{r}^2}{\Bar{T}}}z^{(0)}\right)\mathrm{Li}_{5/2}\left(-e^{-\frac{\Bar{r}^2}{\Bar{T}}}z^{(0)}\right)\rightarrow\frac{64\Bar{r}^2(\Bar{r}^2-1)^4}{5\pi^2},
$
which shows that the integrand vanishes at $\Bar{r}=1$. This is consistent with the LDA, where the density goes to zero at  $r=R_F$. Since  the density of the cloud is zero for $r>R_F$, the upper limit of the integration can be changed from $\Bar{r}=\infty$ to $\Bar{r}=1$, 
\begin{equation}
    \frac{C_1^\mathrm{trap}}{N[\mathrm{Re}(a_1)]^2(k_F^{\mathrm{trap}})^5}\xrightarrow[]{T\rightarrow0}\int_0^1d\Bar{r} \frac{64\Bar{r}^2(\Bar{r}^2-1)^4}{5\pi^2}=\frac{8192}{17325\pi^2}.
\end{equation}

We note that our work does not rule out the existence of other phases in the extremely low-temperature regime. If, e.g., a BCS phase existed, the BCS transition temperature would be exponentially small in the weakly-interacting regime  ($|a_l| k_F^{2l+1} \ll 1$) considered in this work. Evaluating the low-temperature normal-phase behavior would still be justified by considering the limiting $T/T_F \rightarrow 0$ expressions discussed above.


\begin{thebibliography}{71}%
\makeatletter
\providecommand \@ifxundefined [1]{%
 \@ifx{#1\undefined}
}%
\providecommand \@ifnum [1]{%
 \ifnum #1\expandafter \@firstoftwo
 \else \expandafter \@secondoftwo
 \fi
}%
\providecommand \@ifx [1]{%
 \ifx #1\expandafter \@firstoftwo
 \else \expandafter \@secondoftwo
 \fi
}%
\providecommand \natexlab [1]{#1}%
\providecommand \enquote  [1]{#1}%
\providecommand \bibnamefont  [1]{#1}%
\providecommand \bibfnamefont [1]{#1}%
\providecommand \citenamefont [1]{#1}%
\providecommand \href@noop [0]{\@secondoftwo}%
\providecommand \href [0]{\begingroup \@sanitize@url \@href}%
\providecommand \@href[1]{\@@startlink{#1}\@@href}%
\providecommand \@@href[1]{\endgroup#1\@@endlink}%
\providecommand \@sanitize@url [0]{\catcode `\\12\catcode `\$12\catcode `\&12\catcode `\#12\catcode `\^12\catcode `\_12\catcode `\%12\relax}%
\providecommand \@@startlink[1]{}%
\providecommand \@@endlink[0]{}%
\providecommand \url  [0]{\begingroup\@sanitize@url \@url }%
\providecommand \@url [1]{\endgroup\@href {#1}{\urlprefix }}%
\providecommand \urlprefix  [0]{URL }%
\providecommand \Eprint [0]{\href }%
\providecommand \doibase [0]{https://dx.doi.org}%
\providecommand \selectlanguage [0]{\@gobble}%
\providecommand \bibinfo  [0]{\@secondoftwo}%
\providecommand \bibfield  [0]{\@secondoftwo}%
\providecommand \translation [1]{[#1]}%
\providecommand \BibitemOpen [0]{}%
\providecommand \bibitemStop [0]{}%
\providecommand \bibitemNoStop [0]{.\EOS\space}%
\providecommand \EOS [0]{\spacefactor3000\relax}%
\providecommand \BibitemShut  [1]{\csname bibitem#1\endcsname}%
\let\auto@bib@innerbib\@empty
\bibitem [{\citenamefont {Huang}(2008)}]{huang2008statistical}%
  \BibitemOpen
  \bibfield  {author} {\bibinfo {author} {\bibfnamefont {K.}~\bibnamefont {Huang}},\ }\href@noop {} {\emph {\bibinfo {title} {Statistical Mechanics}}}\ (\bibinfo  {publisher} {{John Wiley \& Sons}},\ \bibinfo {address} {New York},\ \bibinfo {year} {2008})\BibitemShut {NoStop}%
\bibitem [{\citenamefont {Bohn}\ \emph {et~al.}(2017)\citenamefont {Bohn}, \citenamefont {Rey},\ and\ \citenamefont {Ye}}]{bohn2017cold}%
  \BibitemOpen
  \bibfield  {author} {\bibinfo {author} {\bibfnamefont {J.~L.}\ \bibnamefont {Bohn}}, \bibinfo {author} {\bibfnamefont {A.~M.}\ \bibnamefont {Rey}}, \ and\ \bibinfo {author} {\bibfnamefont {J.}~\bibnamefont {Ye}},\ }\bibfield  {title} {\bibinfo {title} {Cold molecules: {{Progress}} in quantum engineering of chemistry and quantum matter},\ }\href {\doibase10.1126/science.aam6299} {\bibfield  {journal} {\bibinfo  {journal} {Science}\ }\textbf {\bibinfo {volume} {357}},\ \bibinfo {pages} {1002} (\bibinfo {year} {2017})}\BibitemShut {NoStop}%
\bibitem [{\citenamefont {Anderegg}\ \emph {et~al.}(2018)\citenamefont {Anderegg}, \citenamefont {Augenbraun}, \citenamefont {Bao}, \citenamefont {Burchesky}, \citenamefont {Cheuk}, \citenamefont {Ketterle},\ and\ \citenamefont {Doyle}}]{anderegg2018laser}%
  \BibitemOpen
  \bibfield  {author} {\bibinfo {author} {\bibfnamefont {L.}~\bibnamefont {Anderegg}}, \bibinfo {author} {\bibfnamefont {B.~L.}\ \bibnamefont {Augenbraun}}, \bibinfo {author} {\bibfnamefont {Y.}~\bibnamefont {Bao}}, \bibinfo {author} {\bibfnamefont {S.}~\bibnamefont {Burchesky}}, \bibinfo {author} {\bibfnamefont {L.~W.}\ \bibnamefont {Cheuk}}, \bibinfo {author} {\bibfnamefont {W.}~\bibnamefont {Ketterle}}, \ and\ \bibinfo {author} {\bibfnamefont {J.~M.}\ \bibnamefont {Doyle}},\ }\bibfield  {title} {\bibinfo {title} {Laser cooling of optically trapped molecules},\ }\href {\doibase10.1038/s41567-018-0191-z} {\bibfield  {journal} {\bibinfo  {journal} {Nat. Phys.}\ }\textbf {\bibinfo {volume} {14}},\ \bibinfo {pages} {890} (\bibinfo {year} {2018})}\BibitemShut {NoStop}%
\bibitem [{\citenamefont {Ni}\ \emph {et~al.}(2008)\citenamefont {Ni}, \citenamefont {Ospelkaus}, \citenamefont {De~Miranda}, \citenamefont {Pe'Er}, \citenamefont {Neyenhuis}, \citenamefont {Zirbel}, \citenamefont {Kotochigova}, \citenamefont {Julienne}, \citenamefont {Jin},\ and\ \citenamefont {Ye}}]{ni2008high}%
  \BibitemOpen
  \bibfield  {author} {\bibinfo {author} {\bibfnamefont {K.-K.}\ \bibnamefont {Ni}}, \bibinfo {author} {\bibfnamefont {S.}~\bibnamefont {Ospelkaus}}, \bibinfo {author} {\bibfnamefont {{\relax M. H. G}.}~\bibnamefont {De~Miranda}}, \bibinfo {author} {\bibfnamefont {A.}~\bibnamefont {Pe'Er}}, \bibinfo {author} {\bibfnamefont {B.}~\bibnamefont {Neyenhuis}}, \bibinfo {author} {\bibfnamefont {{\relax J. J}.}~\bibnamefont {Zirbel}}, \bibinfo {author} {\bibfnamefont {S.}~\bibnamefont {Kotochigova}}, \bibinfo {author} {\bibfnamefont {{\relax P. S}.}~\bibnamefont {Julienne}}, \bibinfo {author} {\bibfnamefont {{\relax D. S}.}~\bibnamefont {Jin}}, \ and\ \bibinfo {author} {\bibfnamefont {J.}~\bibnamefont {Ye}},\ }\bibfield  {title} {\bibinfo {title} {{A High Phase-Space-Density Gas of Polar Molecules}},\ }\href {\doibase10.1126/science.1163861} {\bibfield  {journal} {\bibinfo  {journal} {Science}\ }\textbf {\bibinfo {volume} {322}},\ \bibinfo {pages} {231} (\bibinfo {year} {2008})}\BibitemShut {NoStop}%
\bibitem [{\citenamefont {Park}\ \emph {et~al.}(2015)\citenamefont {Park}, \citenamefont {Will},\ and\ \citenamefont {Zwierlein}}]{park2015ultracold}%
  \BibitemOpen
  \bibfield  {author} {\bibinfo {author} {\bibfnamefont {J.~W.}\ \bibnamefont {Park}}, \bibinfo {author} {\bibfnamefont {S.~A.}\ \bibnamefont {Will}}, \ and\ \bibinfo {author} {\bibfnamefont {M.~W.}\ \bibnamefont {Zwierlein}},\ }\bibfield  {title} {\bibinfo {title} {Ultracold {{Dipolar Gas}} of {{Fermionic}} ${}^{23}\mathrm{Na}{}^{40}\mathrm{K}$ {{Molecules}} in {{Their Absolute Ground State}}},\ }\href {\doibase10.1103/PhysRevLett.114.205302} {\bibfield  {journal} {\bibinfo  {journal} {Phys. Rev. Lett.}\ }\textbf {\bibinfo {volume} {114}},\ \bibinfo {pages} {205302} (\bibinfo {year} {2015})}\BibitemShut {NoStop}%
\bibitem [{\citenamefont {See{\ss}elberg}\ \emph {et~al.}(2018)\citenamefont {See{\ss}elberg}, \citenamefont {Buchheim}, \citenamefont {Lu}, \citenamefont {Schneider}, \citenamefont {Luo}, \citenamefont {Tiemann}, \citenamefont {Bloch},\ and\ \citenamefont {Gohle}}]{seesselberg2018modeling}%
  \BibitemOpen
  \bibfield  {author} {\bibinfo {author} {\bibfnamefont {F.}~\bibnamefont {See{\ss}elberg}}, \bibinfo {author} {\bibfnamefont {N.}~\bibnamefont {Buchheim}}, \bibinfo {author} {\bibfnamefont {Z.-K.}\ \bibnamefont {Lu}}, \bibinfo {author} {\bibfnamefont {T.}~\bibnamefont {Schneider}}, \bibinfo {author} {\bibfnamefont {X.-Y.}\ \bibnamefont {Luo}}, \bibinfo {author} {\bibfnamefont {E.}~\bibnamefont {Tiemann}}, \bibinfo {author} {\bibfnamefont {I.}~\bibnamefont {Bloch}}, \ and\ \bibinfo {author} {\bibfnamefont {C.}~\bibnamefont {Gohle}},\ }\bibfield  {title} {\bibinfo {title} {Modeling the adiabatic creation of ultracold polar ${}^{23}\mathrm{Na}{}^{40}\mathrm{K}$ molecules},\ }\href {\doibase10.1103/PhysRevA.97.013405} {\bibfield  {journal} {\bibinfo  {journal} {Phys. Rev. A}\ }\textbf {\bibinfo {volume} {97}},\ \bibinfo {pages} {013405} (\bibinfo {year} {2018})}\BibitemShut {NoStop}%
\bibitem [{\citenamefont {Takekoshi}\ \emph {et~al.}(2014)\citenamefont {Takekoshi}, \citenamefont {Reichs{\"o}llner}, \citenamefont {Schindewolf}, \citenamefont {Hutson}, \citenamefont {Le~Sueur}, \citenamefont {Dulieu}, \citenamefont {Ferlaino}, \citenamefont {Grimm},\ and\ \citenamefont {N{\"a}gerl}}]{takekoshi2014ultracold}%
  \BibitemOpen
  \bibfield  {author} {\bibinfo {author} {\bibfnamefont {T.}~\bibnamefont {Takekoshi}}, \bibinfo {author} {\bibfnamefont {L.}~\bibnamefont {Reichs{\"o}llner}}, \bibinfo {author} {\bibfnamefont {A.}~\bibnamefont {Schindewolf}}, \bibinfo {author} {\bibfnamefont {J.~M.}\ \bibnamefont {Hutson}}, \bibinfo {author} {\bibfnamefont {C.~R.}\ \bibnamefont {Le~Sueur}}, \bibinfo {author} {\bibfnamefont {O.}~\bibnamefont {Dulieu}}, \bibinfo {author} {\bibfnamefont {F.}~\bibnamefont {Ferlaino}}, \bibinfo {author} {\bibfnamefont {R.}~\bibnamefont {Grimm}}, \ and\ \bibinfo {author} {\bibfnamefont {H.-C.}\ \bibnamefont {N{\"a}gerl}},\ }\bibfield  {title} {\bibinfo {title} {Ultracold {{Dense Samples}} of {{Dipolar RbCs Molecules}} in the {{Rovibrational}} and {{Hyperfine Ground State}}},\ }\href {\doibase10.1103/PhysRevLett.113.205301} {\bibfield  {journal} {\bibinfo  {journal} {Phys. Rev. Lett.}\ }\textbf {\bibinfo {volume} {113}},\ \bibinfo {pages} {205301} (\bibinfo {year} {2014})}\BibitemShut {NoStop}%
\bibitem [{\citenamefont {Molony}\ \emph {et~al.}(2014)\citenamefont {Molony}, \citenamefont {Gregory}, \citenamefont {Ji}, \citenamefont {Lu}, \citenamefont {K{\"o}ppinger}, \citenamefont {Le~Sueur}, \citenamefont {Blackley}, \citenamefont {Hutson},\ and\ \citenamefont {Cornish}}]{molony2014creation}%
  \BibitemOpen
  \bibfield  {author} {\bibinfo {author} {\bibfnamefont {P.~K.}\ \bibnamefont {Molony}}, \bibinfo {author} {\bibfnamefont {P.~D.}\ \bibnamefont {Gregory}}, \bibinfo {author} {\bibfnamefont {Z.}~\bibnamefont {Ji}}, \bibinfo {author} {\bibfnamefont {B.}~\bibnamefont {Lu}}, \bibinfo {author} {\bibfnamefont {M.~P.}\ \bibnamefont {K{\"o}ppinger}}, \bibinfo {author} {\bibfnamefont {C.~R.}\ \bibnamefont {Le~Sueur}}, \bibinfo {author} {\bibfnamefont {C.~L.}\ \bibnamefont {Blackley}}, \bibinfo {author} {\bibfnamefont {J.~M.}\ \bibnamefont {Hutson}}, \ and\ \bibinfo {author} {\bibfnamefont {S.~L.}\ \bibnamefont {Cornish}},\ }\bibfield  {title} {\bibinfo {title} {Creation of {{Ultracold}} ${}^{87}\mathrm{Rb}{}^{133}\mathrm{Cs}$ {{Molecules}} in the {{Rovibrational Ground State}}},\ }\href {\doibase10.1103/PhysRevLett.113.255301} {\bibfield  {journal} {\bibinfo  {journal} {Phys. Rev. Lett.}\ }\textbf {\bibinfo {volume} {113}},\ \bibinfo {pages} {255301} (\bibinfo {year} {2014})}\BibitemShut {NoStop}%
\bibitem [{\citenamefont {Guo}\ \emph {et~al.}(2016)\citenamefont {Guo}, \citenamefont {Zhu}, \citenamefont {Lu}, \citenamefont {Ye}, \citenamefont {Wang}, \citenamefont {Vexiau}, \citenamefont {{Bouloufa-Maafa}}, \citenamefont {Qu{\'e}m{\'e}ner}, \citenamefont {Dulieu},\ and\ \citenamefont {Wang}}]{guo2016creation}%
  \BibitemOpen
  \bibfield  {author} {\bibinfo {author} {\bibfnamefont {M.}~\bibnamefont {Guo}}, \bibinfo {author} {\bibfnamefont {B.}~\bibnamefont {Zhu}}, \bibinfo {author} {\bibfnamefont {B.}~\bibnamefont {Lu}}, \bibinfo {author} {\bibfnamefont {X.}~\bibnamefont {Ye}}, \bibinfo {author} {\bibfnamefont {F.}~\bibnamefont {Wang}}, \bibinfo {author} {\bibfnamefont {R.}~\bibnamefont {Vexiau}}, \bibinfo {author} {\bibfnamefont {N.}~\bibnamefont {{Bouloufa-Maafa}}}, \bibinfo {author} {\bibfnamefont {G.}~\bibnamefont {Qu{\'e}m{\'e}ner}}, \bibinfo {author} {\bibfnamefont {O.}~\bibnamefont {Dulieu}}, \ and\ \bibinfo {author} {\bibfnamefont {D.}~\bibnamefont {Wang}},\ }\bibfield  {title} {\bibinfo {title} {Creation of an {{Ultracold Gas}} of {{Ground-State Dipolar}} ${}^{23}\mathrm{Na}{}^{87}\mathrm{Rb}$ {{Molecules}}},\ }\href {\doibase10.1103/PhysRevLett.116.205303} {\bibfield  {journal} {\bibinfo  {journal} {Phys. Rev. Lett.}\ }\textbf {\bibinfo {volume} {116}},\ \bibinfo {pages} {205303} (\bibinfo {year} {2016})}\BibitemShut
  {NoStop}%
\bibitem [{\citenamefont {Rvachov}\ \emph {et~al.}(2017)\citenamefont {Rvachov}, \citenamefont {Son}, \citenamefont {Sommer}, \citenamefont {Ebadi}, \citenamefont {Park}, \citenamefont {Zwierlein}, \citenamefont {Ketterle},\ and\ \citenamefont {Jamison}}]{rvachov2017longlived}%
  \BibitemOpen
  \bibfield  {author} {\bibinfo {author} {\bibfnamefont {T.~M.}\ \bibnamefont {Rvachov}}, \bibinfo {author} {\bibfnamefont {H.}~\bibnamefont {Son}}, \bibinfo {author} {\bibfnamefont {A.~T.}\ \bibnamefont {Sommer}}, \bibinfo {author} {\bibfnamefont {S.}~\bibnamefont {Ebadi}}, \bibinfo {author} {\bibfnamefont {J.~J.}\ \bibnamefont {Park}}, \bibinfo {author} {\bibfnamefont {M.~W.}\ \bibnamefont {Zwierlein}}, \bibinfo {author} {\bibfnamefont {W.}~\bibnamefont {Ketterle}}, \ and\ \bibinfo {author} {\bibfnamefont {A.~O.}\ \bibnamefont {Jamison}},\ }\bibfield  {title} {\bibinfo {title} {Long-{{Lived Ultracold Molecules}} with {{Electric}} and {{Magnetic Dipole Moments}}},\ }\href {\doibase10.1103/PhysRevLett.119.143001} {\bibfield  {journal} {\bibinfo  {journal} {Phys. Rev. Lett.}\ }\textbf {\bibinfo {volume} {119}},\ \bibinfo {pages} {143001} (\bibinfo {year} {2017})}\BibitemShut {NoStop}%
\bibitem [{\citenamefont {Duda}\ \emph {et~al.}(2023)\citenamefont {Duda}, \citenamefont {Chen}, \citenamefont {Bause}, \citenamefont {Schindewolf}, \citenamefont {Bloch},\ and\ \citenamefont {Luo}}]{duda2023longlived}%
  \BibitemOpen
  \bibfield  {author} {\bibinfo {author} {\bibfnamefont {M.}~\bibnamefont {Duda}}, \bibinfo {author} {\bibfnamefont {X.-Y.}\ \bibnamefont {Chen}}, \bibinfo {author} {\bibfnamefont {R.}~\bibnamefont {Bause}}, \bibinfo {author} {\bibfnamefont {A.}~\bibnamefont {Schindewolf}}, \bibinfo {author} {\bibfnamefont {I.}~\bibnamefont {Bloch}}, \ and\ \bibinfo {author} {\bibfnamefont {X.-Y.}\ \bibnamefont {Luo}},\ }\bibfield  {title} {\bibinfo {title} {{Long-Lived Fermionic Feshbach Molecules with Tunable $p$-Wave Interactions}},\ }\href {\doibase10.1103/PhysRevA.107.053322} {\bibfield  {journal} {\bibinfo  {journal} {Phys. Rev. A}\ }\textbf {\bibinfo {volume} {107}},\ \bibinfo {pages} {053322} (\bibinfo {year} {2023})}\BibitemShut {NoStop}%
\bibitem [{\citenamefont {Tan}(2008{\natexlab{a}})}]{tan2008energetics}%
  \BibitemOpen
  \bibfield  {author} {\bibinfo {author} {\bibfnamefont {S.}~\bibnamefont {Tan}},\ }\bibfield  {title} {\bibinfo {title} {Energetics of a strongly correlated {{Fermi}} gas},\ }\href {\doibase10.1016/j.aop.2008.03.004} {\bibfield  {journal} {\bibinfo  {journal} {Ann. Phys.}\ }\textbf {\bibinfo {volume} {323}},\ \bibinfo {pages} {2952} (\bibinfo {year} {2008}{\natexlab{a}})}\BibitemShut {NoStop}%
\bibitem [{\citenamefont {Tan}(2008{\natexlab{b}})}]{tan2008generalized}%
  \BibitemOpen
  \bibfield  {author} {\bibinfo {author} {\bibfnamefont {S.}~\bibnamefont {Tan}},\ }\bibfield  {title} {\bibinfo {title} {Generalized virial theorem and pressure relation for a strongly correlated {{Fermi}} gas},\ }\href {\doibase10.1016/j.aop.2008.03.003} {\bibfield  {journal} {\bibinfo  {journal} {Ann. Phys.}\ }\textbf {\bibinfo {volume} {323}},\ \bibinfo {pages} {2987} (\bibinfo {year} {2008}{\natexlab{b}})}\BibitemShut {NoStop}%
\bibitem [{\citenamefont {Tan}(2008{\natexlab{c}})}]{tan2008large}%
  \BibitemOpen
  \bibfield  {author} {\bibinfo {author} {\bibfnamefont {S.}~\bibnamefont {Tan}},\ }\bibfield  {title} {\bibinfo {title} {Large momentum part of a strongly correlated {{Fermi}} gas},\ }\href {\doibase10.1016/j.aop.2008.03.005} {\bibfield  {journal} {\bibinfo  {journal} {Ann. Phys.}\ }\textbf {\bibinfo {volume} {323}},\ \bibinfo {pages} {2971} (\bibinfo {year} {2008}{\natexlab{c}})}\BibitemShut {NoStop}%
\bibitem [{\citenamefont {Stewart}\ \emph {et~al.}(2010)\citenamefont {Stewart}, \citenamefont {Gaebler}, \citenamefont {Drake},\ and\ \citenamefont {Jin}}]{stewart2010verification}%
  \BibitemOpen
  \bibfield  {author} {\bibinfo {author} {\bibfnamefont {{\relax D. S}.}~\bibnamefont {Stewart}}, \bibinfo {author} {\bibfnamefont {{\relax J. P}.}~\bibnamefont {Gaebler}}, \bibinfo {author} {\bibfnamefont {{\relax T. E}.}~\bibnamefont {Drake}}, \ and\ \bibinfo {author} {\bibfnamefont {{\relax D. S}.}~\bibnamefont {Jin}},\ }\bibfield  {title} {\bibinfo {title} {Verification of universal relations in a strongly interacting {{Fermi}} gas},\ }\href {\doibase10.1103/PhysRevLett.104.235301} {\bibfield  {journal} {\bibinfo  {journal} {Phys. Rev. Lett.}\ }\textbf {\bibinfo {volume} {104}},\ \bibinfo {pages} {235301} (\bibinfo {year} {2010})}\BibitemShut {NoStop}%
\bibitem [{\citenamefont {Kuhnle}\ \emph {et~al.}(2010)\citenamefont {Kuhnle}, \citenamefont {Hu}, \citenamefont {Liu}, \citenamefont {Dyke}, \citenamefont {Mark}, \citenamefont {Drummond}, \citenamefont {Hannaford},\ and\ \citenamefont {Vale}}]{kuhnle2010universal}%
  \BibitemOpen
  \bibfield  {author} {\bibinfo {author} {\bibfnamefont {{\relax E. D}.}~\bibnamefont {Kuhnle}}, \bibinfo {author} {\bibfnamefont {H.}~\bibnamefont {Hu}}, \bibinfo {author} {\bibfnamefont {X.-J.}\ \bibnamefont {Liu}}, \bibinfo {author} {\bibfnamefont {P.}~\bibnamefont {Dyke}}, \bibinfo {author} {\bibfnamefont {M.}~\bibnamefont {Mark}}, \bibinfo {author} {\bibfnamefont {{\relax P. D}.}~\bibnamefont {Drummond}}, \bibinfo {author} {\bibfnamefont {P.}~\bibnamefont {Hannaford}}, \ and\ \bibinfo {author} {\bibfnamefont {{\relax C. J}.}~\bibnamefont {Vale}},\ }\bibfield  {title} {\bibinfo {title} {Universal behavior of pair correlations in a strongly interacting {{Fermi}} gas},\ }\href {\doibase10.1103/PhysRevLett.105.070402} {\bibfield  {journal} {\bibinfo  {journal} {Phys. Rev. Lett.}\ }\textbf {\bibinfo {volume} {105}},\ \bibinfo {pages} {070402} (\bibinfo {year} {2010})}\BibitemShut {NoStop}%
\bibitem [{\citenamefont {Sagi}\ \emph {et~al.}(2012)\citenamefont {Sagi}, \citenamefont {Drake}, \citenamefont {Paudel},\ and\ \citenamefont {Jin}}]{sagi2012measurementa}%
  \BibitemOpen
  \bibfield  {author} {\bibinfo {author} {\bibfnamefont {Y.}~\bibnamefont {Sagi}}, \bibinfo {author} {\bibfnamefont {T.~E.}\ \bibnamefont {Drake}}, \bibinfo {author} {\bibfnamefont {R.}~\bibnamefont {Paudel}}, \ and\ \bibinfo {author} {\bibfnamefont {D.~S.}\ \bibnamefont {Jin}},\ }\bibfield  {title} {\bibinfo {title} {Measurement of the {{Homogeneous Contact}} of a {{Unitary Fermi Gas}}},\ }\href {\doibase10.1103/PhysRevLett.109.220402} {\bibfield  {journal} {\bibinfo  {journal} {Phys. Rev. Lett.}\ }\textbf {\bibinfo {volume} {109}},\ \bibinfo {pages} {220402} (\bibinfo {year} {2012})}\BibitemShut {NoStop}%
\bibitem [{\citenamefont {Werner}\ and\ \citenamefont {Castin}(2012{\natexlab{a}})}]{werner2012generala}%
  \BibitemOpen
  \bibfield  {author} {\bibinfo {author} {\bibfnamefont {F.}~\bibnamefont {Werner}}\ and\ \bibinfo {author} {\bibfnamefont {Y.}~\bibnamefont {Castin}},\ }\bibfield  {title} {\bibinfo {title} {General relations for quantum gases in two and three dimensions: {{Two-component}} fermions},\ }\href {\doibase10.1103/PhysRevA.86.013626} {\bibfield  {journal} {\bibinfo  {journal} {Phys. Rev. A}\ }\textbf {\bibinfo {volume} {86}},\ \bibinfo {pages} {013626} (\bibinfo {year} {2012}{\natexlab{a}})}\BibitemShut {NoStop}%
\bibitem [{\citenamefont {Wild}\ \emph {et~al.}(2012)\citenamefont {Wild}, \citenamefont {Makotyn}, \citenamefont {Pino}, \citenamefont {Cornell},\ and\ \citenamefont {Jin}}]{wild2012measurements}%
  \BibitemOpen
  \bibfield  {author} {\bibinfo {author} {\bibfnamefont {{\relax R. J}.}~\bibnamefont {Wild}}, \bibinfo {author} {\bibfnamefont {P.}~\bibnamefont {Makotyn}}, \bibinfo {author} {\bibfnamefont {{\relax J. M}.}~\bibnamefont {Pino}}, \bibinfo {author} {\bibfnamefont {{\relax E. A}.}~\bibnamefont {Cornell}}, \ and\ \bibinfo {author} {\bibfnamefont {{\relax D. S}.}~\bibnamefont {Jin}},\ }\bibfield  {title} {\bibinfo {title} {Measurements of {{Tan}}'s contact in an atomic {{Bose-Einstein}} condensate},\ }\href {\doibase10.1103/PhysRevLett.108.145305} {\bibfield  {journal} {\bibinfo  {journal} {Phys. Rev. Lett.}\ }\textbf {\bibinfo {volume} {108}},\ \bibinfo {pages} {145305} (\bibinfo {year} {2012})}\BibitemShut {NoStop}%
\bibitem [{\citenamefont {Hoinka}\ \emph {et~al.}(2013)\citenamefont {Hoinka}, \citenamefont {Lingham}, \citenamefont {Fenech}, \citenamefont {Hu}, \citenamefont {Vale}, \citenamefont {Drut},\ and\ \citenamefont {Gandolfi}}]{hoinka2013precise}%
  \BibitemOpen
  \bibfield  {author} {\bibinfo {author} {\bibfnamefont {S.}~\bibnamefont {Hoinka}}, \bibinfo {author} {\bibfnamefont {M.}~\bibnamefont {Lingham}}, \bibinfo {author} {\bibfnamefont {K.}~\bibnamefont {Fenech}}, \bibinfo {author} {\bibfnamefont {H.}~\bibnamefont {Hu}}, \bibinfo {author} {\bibfnamefont {C.~J.}\ \bibnamefont {Vale}}, \bibinfo {author} {\bibfnamefont {J.~E.}\ \bibnamefont {Drut}}, \ and\ \bibinfo {author} {\bibfnamefont {S.}~\bibnamefont {Gandolfi}},\ }\bibfield  {title} {\bibinfo {title} {Precise determination of the structure factor and contact in a unitary {{Fermi}} gas},\ }\href {\doibase10.1103/PhysRevLett.110.055305} {\bibfield  {journal} {\bibinfo  {journal} {Phys. Rev. Lett.}\ }\textbf {\bibinfo {volume} {110}},\ \bibinfo {pages} {055305} (\bibinfo {year} {2013})}\BibitemShut {NoStop}%
\bibitem [{\citenamefont {Shashi}\ \emph {et~al.}(2014)\citenamefont {Shashi}, \citenamefont {Grusdt}, \citenamefont {Abanin},\ and\ \citenamefont {Demler}}]{shashi2014radiofrequency}%
  \BibitemOpen
  \bibfield  {author} {\bibinfo {author} {\bibfnamefont {A.}~\bibnamefont {Shashi}}, \bibinfo {author} {\bibfnamefont {F.}~\bibnamefont {Grusdt}}, \bibinfo {author} {\bibfnamefont {D.~A.}\ \bibnamefont {Abanin}}, \ and\ \bibinfo {author} {\bibfnamefont {E.}~\bibnamefont {Demler}},\ }\bibfield  {title} {\bibinfo {title} {Radio-frequency spectroscopy of polarons in ultracold {{Bose}} gases},\ }\href {\doibase10.1103/PhysRevA.89.053617} {\bibfield  {journal} {\bibinfo  {journal} {Phys. Rev. A}\ }\textbf {\bibinfo {volume} {89}},\ \bibinfo {pages} {053617} (\bibinfo {year} {2014})}\BibitemShut {NoStop}%
\bibitem [{\citenamefont {Werner}\ and\ \citenamefont {Castin}(2012{\natexlab{b}})}]{werner2012generalb}%
  \BibitemOpen
  \bibfield  {author} {\bibinfo {author} {\bibfnamefont {F.}~\bibnamefont {Werner}}\ and\ \bibinfo {author} {\bibfnamefont {Y.}~\bibnamefont {Castin}},\ }\bibfield  {title} {\bibinfo {title} {General relations for quantum gases in two and three dimensions. {{II}}. {{Bosons}} and mixtures},\ }\href {\doibase10.1103/PhysRevA.86.053633} {\bibfield  {journal} {\bibinfo  {journal} {Phys. Rev. A}\ }\textbf {\bibinfo {volume} {86}},\ \bibinfo {pages} {053633} (\bibinfo {year} {2012}{\natexlab{b}})}\BibitemShut {NoStop}%
\bibitem [{\citenamefont {Yu}\ \emph {et~al.}(2015)\citenamefont {Yu}, \citenamefont {Thywissen},\ and\ \citenamefont {Zhang}}]{yu2015universal}%
  \BibitemOpen
  \bibfield  {author} {\bibinfo {author} {\bibfnamefont {Z.}~\bibnamefont {Yu}}, \bibinfo {author} {\bibfnamefont {J.~H.}\ \bibnamefont {Thywissen}}, \ and\ \bibinfo {author} {\bibfnamefont {S.}~\bibnamefont {Zhang}},\ }\bibfield  {title} {\bibinfo {title} {Universal relations for a {{Fermi}} gas close to a $p$-wave interaction resonance},\ }\href {\doibase10.1103/PhysRevLett.115.135304} {\bibfield  {journal} {\bibinfo  {journal} {Phys. Rev. Lett.}\ }\textbf {\bibinfo {volume} {115}},\ \bibinfo {pages} {135304} (\bibinfo {year} {2015})}\BibitemShut {NoStop}%
\bibitem [{\citenamefont {Luciuk}\ \emph {et~al.}(2016)\citenamefont {Luciuk}, \citenamefont {Trotzky}, \citenamefont {Smale}, \citenamefont {Yu}, \citenamefont {Zhang},\ and\ \citenamefont {Thywissen}}]{luciuk2016evidence}%
  \BibitemOpen
  \bibfield  {author} {\bibinfo {author} {\bibfnamefont {C.}~\bibnamefont {Luciuk}}, \bibinfo {author} {\bibfnamefont {S.}~\bibnamefont {Trotzky}}, \bibinfo {author} {\bibfnamefont {S.}~\bibnamefont {Smale}}, \bibinfo {author} {\bibfnamefont {Z.}~\bibnamefont {Yu}}, \bibinfo {author} {\bibfnamefont {S.}~\bibnamefont {Zhang}}, \ and\ \bibinfo {author} {\bibfnamefont {J.~H.}\ \bibnamefont {Thywissen}},\ }\bibfield  {title} {\bibinfo {title} {Evidence for universal relations describing a gas with $p$-wave interactions},\ }\href {\doibase10.1038/nphys3670} {\bibfield  {journal} {\bibinfo  {journal} {Nat. Phys.}\ }\textbf {\bibinfo {volume} {12}},\ \bibinfo {pages} {599} (\bibinfo {year} {2016})}\BibitemShut {NoStop}%
\bibitem [{\citenamefont {Yoshida}\ and\ \citenamefont {Ueda}(2015)}]{yoshida2015universal}%
  \BibitemOpen
  \bibfield  {author} {\bibinfo {author} {\bibfnamefont {S.~M.}\ \bibnamefont {Yoshida}}\ and\ \bibinfo {author} {\bibfnamefont {M.}~\bibnamefont {Ueda}},\ }\bibfield  {title} {\bibinfo {title} {Universal high-momentum asymptote and thermodynamic relations in a spinless {{Fermi}} gas with a resonant $p$-wave interaction},\ }\href {\doibase10.1103/PhysRevLett.115.135303} {\bibfield  {journal} {\bibinfo  {journal} {Phys. Rev. Lett.}\ }\textbf {\bibinfo {volume} {115}},\ \bibinfo {pages} {135303} (\bibinfo {year} {2015})}\BibitemShut {NoStop}%
\bibitem [{\citenamefont {He}\ \emph {et~al.}(2016)\citenamefont {He}, \citenamefont {Zhang}, \citenamefont {Chan},\ and\ \citenamefont {Zhou}}]{he2016concept}%
  \BibitemOpen
  \bibfield  {author} {\bibinfo {author} {\bibfnamefont {M.}~\bibnamefont {He}}, \bibinfo {author} {\bibfnamefont {S.}~\bibnamefont {Zhang}}, \bibinfo {author} {\bibfnamefont {H.~M.}\ \bibnamefont {Chan}}, \ and\ \bibinfo {author} {\bibfnamefont {Q.}~\bibnamefont {Zhou}},\ }\bibfield  {title} {\bibinfo {title} {Concept of a {{Contact Spectrum}} and {{Its Applications}} in {{Atomic Quantum Hall States}}},\ }\href {\doibase10.1103/PhysRevLett.116.045301} {\bibfield  {journal} {\bibinfo  {journal} {Phys. Rev. Lett.}\ }\textbf {\bibinfo {volume} {116}},\ \bibinfo {pages} {045301} (\bibinfo {year} {2016})}\BibitemShut {NoStop}%
\bibitem [{\citenamefont {Zhang}\ \emph {et~al.}(2017)\citenamefont {Zhang}, \citenamefont {He},\ and\ \citenamefont {Zhou}}]{zhang2017contact}%
  \BibitemOpen
  \bibfield  {author} {\bibinfo {author} {\bibfnamefont {S.-L.}\ \bibnamefont {Zhang}}, \bibinfo {author} {\bibfnamefont {M.}~\bibnamefont {He}}, \ and\ \bibinfo {author} {\bibfnamefont {Q.}~\bibnamefont {Zhou}},\ }\bibfield  {title} {\bibinfo {title} {Contact matrix in dilute quantum systems},\ }\href {\doibase10.1103/PhysRevA.95.062702} {\bibfield  {journal} {\bibinfo  {journal} {Phys. Rev. A}\ }\textbf {\bibinfo {volume} {95}},\ \bibinfo {pages} {062702} (\bibinfo {year} {2017})}\BibitemShut {NoStop}%
\bibitem [{\citenamefont {Song}\ \emph {et~al.}(2020)\citenamefont {Song}, \citenamefont {Yan}, \citenamefont {He}, \citenamefont {Ren}, \citenamefont {Zhou},\ and\ \citenamefont {Jo}}]{song2020evidence}%
  \BibitemOpen
  \bibfield  {author} {\bibinfo {author} {\bibfnamefont {B.}~\bibnamefont {Song}}, \bibinfo {author} {\bibfnamefont {Y.}~\bibnamefont {Yan}}, \bibinfo {author} {\bibfnamefont {C.}~\bibnamefont {He}}, \bibinfo {author} {\bibfnamefont {Z.}~\bibnamefont {Ren}}, \bibinfo {author} {\bibfnamefont {Q.}~\bibnamefont {Zhou}}, \ and\ \bibinfo {author} {\bibfnamefont {G.-B.}\ \bibnamefont {Jo}},\ }\bibfield  {title} {\bibinfo {title} {Evidence for {{Bosonization}} in a {{Three-Dimensional Gas}} of {$\mathrm{SU}(N)$} {{Fermions}}},\ }\href {\doibase10.1103/PhysRevX.10.041053} {\bibfield  {journal} {\bibinfo  {journal} {Phys. Rev. X}\ }\textbf {\bibinfo {volume} {10}},\ \bibinfo {pages} {041053} (\bibinfo {year} {2020})}\BibitemShut {NoStop}%
\bibitem [{\citenamefont {Braaten}\ and\ \citenamefont {Platter}(2008)}]{braaten2008exact}%
  \BibitemOpen
  \bibfield  {author} {\bibinfo {author} {\bibfnamefont {E.}~\bibnamefont {Braaten}}\ and\ \bibinfo {author} {\bibfnamefont {L.}~\bibnamefont {Platter}},\ }\bibfield  {title} {\bibinfo {title} {Exact relations for a strongly interacting {{Fermi}} gas from the operator product expansion},\ }\href {\doibase10.1103/PhysRevLett.100.205301} {\bibfield  {journal} {\bibinfo  {journal} {Phys. Rev. Lett.}\ }\textbf {\bibinfo {volume} {100}},\ \bibinfo {pages} {205301} (\bibinfo {year} {2008})}\BibitemShut {NoStop}%
\bibitem [{\citenamefont {Braaten}\ and\ \citenamefont {Hammer}(2013)}]{braaten2013universal}%
  \BibitemOpen
  \bibfield  {author} {\bibinfo {author} {\bibfnamefont {E.}~\bibnamefont {Braaten}}\ and\ \bibinfo {author} {\bibfnamefont {H.~W.}\ \bibnamefont {Hammer}},\ }\bibfield  {title} {\bibinfo {title} {Universal relation for the inelastic two-body loss rate},\ }\href {\doibase10.1088/0953-4075/46/21/215203} {\bibfield  {journal} {\bibinfo  {journal} {J. Phys. B}\ }\textbf {\bibinfo {volume} {46}},\ \bibinfo {pages} {215203} (\bibinfo {year} {2013})}\BibitemShut {NoStop}%
\bibitem [{\citenamefont {Braaten}\ \emph {et~al.}(2017)\citenamefont {Braaten}, \citenamefont {Hammer},\ and\ \citenamefont {Lepage}}]{braaten2017lindblad}%
  \BibitemOpen
  \bibfield  {author} {\bibinfo {author} {\bibfnamefont {E.}~\bibnamefont {Braaten}}, \bibinfo {author} {\bibfnamefont {H.-W.}\ \bibnamefont {Hammer}}, \ and\ \bibinfo {author} {\bibfnamefont {G.~P.}\ \bibnamefont {Lepage}},\ }\bibfield  {title} {\bibinfo {title} {Lindblad equation for the inelastic loss of ultracold atoms},\ }\href {\doibase10.1103/PhysRevA.95.012708} {\bibfield  {journal} {\bibinfo  {journal} {Phys. Rev. A}\ }\textbf {\bibinfo {volume} {95}},\ \bibinfo {pages} {012708} (\bibinfo {year} {2017})}\BibitemShut {NoStop}%
\bibitem [{\citenamefont {He}\ \emph {et~al.}(2020)\citenamefont {He}, \citenamefont {Lv}, \citenamefont {Lin},\ and\ \citenamefont {Zhou}}]{he2020universal}%
  \BibitemOpen
  \bibfield  {author} {\bibinfo {author} {\bibfnamefont {M.}~\bibnamefont {He}}, \bibinfo {author} {\bibfnamefont {C.}~\bibnamefont {Lv}}, \bibinfo {author} {\bibfnamefont {H.-Q.}\ \bibnamefont {Lin}}, \ and\ \bibinfo {author} {\bibfnamefont {Q.}~\bibnamefont {Zhou}},\ }\bibfield  {title} {\bibinfo {title} {Universal relations for ultracold reactive molecules},\ }\href {\doibase10.1126/sciadv.abd4699} {\bibfield  {journal} {\bibinfo  {journal} {Sci. Adv.}\ }\textbf {\bibinfo {volume} {6}},\ \bibinfo {pages} {eabd4699} (\bibinfo {year} {2020})}\BibitemShut {NoStop}%
\bibitem [{\citenamefont {Gao}\ \emph {et~al.}(2023)\citenamefont {Gao}, \citenamefont {Blume},\ and\ \citenamefont {Yan}}]{gao2023temperaturedependent}%
  \BibitemOpen
  \bibfield  {author} {\bibinfo {author} {\bibfnamefont {X.-Y.}\ \bibnamefont {Gao}}, \bibinfo {author} {\bibfnamefont {D.}~\bibnamefont {Blume}}, \ and\ \bibinfo {author} {\bibfnamefont {Y.}~\bibnamefont {Yan}},\ }\bibfield  {title} {\bibinfo {title} {Temperature-{{Dependent Contact}} of {{Weakly Interacting Single-Component Fermi Gases}} and {{Loss Rate}} of {{Degenerate Polar Molecules}}},\ }\href {\doibase10.1103/PhysRevLett.131.043401} {\bibfield  {journal} {\bibinfo  {journal} {Phys. Rev. Lett.}\ }\textbf {\bibinfo {volume} {131}},\ \bibinfo {pages} {043401} (\bibinfo {year} {2023})}\BibitemShut {NoStop}%
\bibitem [{\citenamefont {Andersen}(2004)}]{andersen2004theory}%
  \BibitemOpen
  \bibfield  {author} {\bibinfo {author} {\bibfnamefont {J.~O.}\ \bibnamefont {Andersen}},\ }\bibfield  {title} {\bibinfo {title} {Theory of the weakly interacting {{Bose}} gas},\ }\href {\doibase10.1103/RevModPhys.76.599} {\bibfield  {journal} {\bibinfo  {journal} {Rev. Mod. Phys.}\ }\textbf {\bibinfo {volume} {76}},\ \bibinfo {pages} {599} (\bibinfo {year} {2004})}\BibitemShut {NoStop}%
\bibitem [{\citenamefont {Ding}\ and\ \citenamefont {Zhang}(2019)}]{ding2019fermiliquid}%
  \BibitemOpen
  \bibfield  {author} {\bibinfo {author} {\bibfnamefont {S.}~\bibnamefont {Ding}}\ and\ \bibinfo {author} {\bibfnamefont {S.}~\bibnamefont {Zhang}},\ }\bibfield  {title} {\bibinfo {title} {Fermi-{{Liquid Description}} of a {{Single-Component Fermi Gas}} with $p$-{{Wave Interactions}}},\ }\href {\doibase10.1103/PhysRevLett.123.070404} {\bibfield  {journal} {\bibinfo  {journal} {Phys. Rev. Lett.}\ }\textbf {\bibinfo {volume} {123}},\ \bibinfo {pages} {070404} (\bibinfo {year} {2019})}\BibitemShut {NoStop}%
\bibitem [{\citenamefont {Yao}\ \emph {et~al.}(2019)\citenamefont {Yao}, \citenamefont {Qi}, \citenamefont {Liu}, \citenamefont {Wang}, \citenamefont {Wang}, \citenamefont {Wu}, \citenamefont {Chen}, \citenamefont {Zhang}, \citenamefont {Zhai}, \citenamefont {Chen},\ and\ \citenamefont {Pan}}]{yao2019degenerate}%
  \BibitemOpen
  \bibfield  {author} {\bibinfo {author} {\bibfnamefont {X.-C.}\ \bibnamefont {Yao}}, \bibinfo {author} {\bibfnamefont {R.}~\bibnamefont {Qi}}, \bibinfo {author} {\bibfnamefont {X.-P.}\ \bibnamefont {Liu}}, \bibinfo {author} {\bibfnamefont {X.-Q.}\ \bibnamefont {Wang}}, \bibinfo {author} {\bibfnamefont {Y.-X.}\ \bibnamefont {Wang}}, \bibinfo {author} {\bibfnamefont {Y.-P.}\ \bibnamefont {Wu}}, \bibinfo {author} {\bibfnamefont {H.-Z.}\ \bibnamefont {Chen}}, \bibinfo {author} {\bibfnamefont {P.}~\bibnamefont {Zhang}}, \bibinfo {author} {\bibfnamefont {H.}~\bibnamefont {Zhai}}, \bibinfo {author} {\bibfnamefont {Y.-A.}\ \bibnamefont {Chen}}, \ and\ \bibinfo {author} {\bibfnamefont {J.-W.}\ \bibnamefont {Pan}},\ }\bibfield  {title} {\bibinfo {title} {{Degenerate Bose Gases near a $d$-Wave Shape Resonance}},\ }\href {\doibase10.1038/s41567-019-0455-2} {\bibfield  {journal} {\bibinfo  {journal} {Nat. Phys.}\ }\textbf {\bibinfo {volume} {15}},\ \bibinfo {pages} {570} (\bibinfo {year} {2019})}\BibitemShut {NoStop}%
\bibitem [{\citenamefont {Cui}\ \emph {et~al.}(2017)\citenamefont {Cui}, \citenamefont {Shen}, \citenamefont {Deng}, \citenamefont {Dong}, \citenamefont {Chen}, \citenamefont {L{\"u}}, \citenamefont {Gao}, \citenamefont {Tey},\ and\ \citenamefont {You}}]{cui2017observation}%
  \BibitemOpen
  \bibfield  {author} {\bibinfo {author} {\bibfnamefont {Y.}~\bibnamefont {Cui}}, \bibinfo {author} {\bibfnamefont {C.}~\bibnamefont {Shen}}, \bibinfo {author} {\bibfnamefont {M.}~\bibnamefont {Deng}}, \bibinfo {author} {\bibfnamefont {S.}~\bibnamefont {Dong}}, \bibinfo {author} {\bibfnamefont {C.}~\bibnamefont {Chen}}, \bibinfo {author} {\bibfnamefont {R.}~\bibnamefont {L{\"u}}}, \bibinfo {author} {\bibfnamefont {B.}~\bibnamefont {Gao}}, \bibinfo {author} {\bibfnamefont {M.~K.}\ \bibnamefont {Tey}}, \ and\ \bibinfo {author} {\bibfnamefont {L.}~\bibnamefont {You}},\ }\bibfield  {title} {\bibinfo {title} {Observation of {{Broad}} $d$-{{Wave Feshbach Resonances}} with a {{Triplet Structure}}},\ }\href {\doibase10.1103/PhysRevLett.119.203402} {\bibfield  {journal} {\bibinfo  {journal} {Phys. Rev. Lett.}\ }\textbf {\bibinfo {volume} {119}},\ \bibinfo {pages} {203402} (\bibinfo {year} {2017})}\BibitemShut {NoStop}%
\bibitem [{\citenamefont {Liu}\ \emph {et~al.}(2018)\citenamefont {Liu}, \citenamefont {Yao}, \citenamefont {Qi}, \citenamefont {Wang}, \citenamefont {Wang}, \citenamefont {Chen},\ and\ \citenamefont {Pan}}]{liu2018feshbach}%
  \BibitemOpen
  \bibfield  {author} {\bibinfo {author} {\bibfnamefont {X.-P.}\ \bibnamefont {Liu}}, \bibinfo {author} {\bibfnamefont {X.-C.}\ \bibnamefont {Yao}}, \bibinfo {author} {\bibfnamefont {R.}~\bibnamefont {Qi}}, \bibinfo {author} {\bibfnamefont {X.-Q.}\ \bibnamefont {Wang}}, \bibinfo {author} {\bibfnamefont {Y.-X.}\ \bibnamefont {Wang}}, \bibinfo {author} {\bibfnamefont {Y.-A.}\ \bibnamefont {Chen}}, \ and\ \bibinfo {author} {\bibfnamefont {J.-W.}\ \bibnamefont {Pan}},\ }\bibfield  {title} {\bibinfo {title} {Feshbach spectroscopy of an ultracold ${}^{41}\mathrm{K}$-${}^{6}\mathrm{Li}$ mixture and ${}^{41}\mathrm{K}$ atoms},\ }\href {\doibase10.1103/PhysRevA.98.022704} {\bibfield  {journal} {\bibinfo  {journal} {Phys. Rev. A}\ }\textbf {\bibinfo {volume} {98}},\ \bibinfo {pages} {022704} (\bibinfo {year} {2018})}\BibitemShut {NoStop}%
\bibitem [{\citenamefont {Shi}\ \emph {et~al.}(2023)\citenamefont {Shi}, \citenamefont {Li}, \citenamefont {Wang}, \citenamefont {Han}, \citenamefont {Huang}, \citenamefont {Meng}, \citenamefont {Chen},\ and\ \citenamefont {Zhang}}]{shi2023collective}%
  \BibitemOpen
  \bibfield  {author} {\bibinfo {author} {\bibfnamefont {Z.}~\bibnamefont {Shi}}, \bibinfo {author} {\bibfnamefont {Z.}~\bibnamefont {Li}}, \bibinfo {author} {\bibfnamefont {P.}~\bibnamefont {Wang}}, \bibinfo {author} {\bibfnamefont {W.}~\bibnamefont {Han}}, \bibinfo {author} {\bibfnamefont {L.}~\bibnamefont {Huang}}, \bibinfo {author} {\bibfnamefont {Z.}~\bibnamefont {Meng}}, \bibinfo {author} {\bibfnamefont {L.}~\bibnamefont {Chen}}, \ and\ \bibinfo {author} {\bibfnamefont {J.}~\bibnamefont {Zhang}},\ }\bibfield  {title} {\bibinfo {title} {{Collective Excitation of {{Bose}}{\textendash}Einstein Condensate of ${}^{23}$Na via High-Partial Wave Feshbach Resonance}},\ }\href {\doibase10.1088/1367-2630/acbd67} {\bibfield  {journal} {\bibinfo  {journal} {New J. Phys.}\ }\textbf {\bibinfo {volume} {25}},\ \bibinfo {pages} {023032} (\bibinfo {year} {2023})}\BibitemShut {NoStop}%
\bibitem [{\citenamefont {Ho}\ and\ \citenamefont {Diener}(2005)}]{ho2005fermion}%
  \BibitemOpen
  \bibfield  {author} {\bibinfo {author} {\bibfnamefont {T.-L.}\ \bibnamefont {Ho}}\ and\ \bibinfo {author} {\bibfnamefont {R.~B.}\ \bibnamefont {Diener}},\ }\bibfield  {title} {\bibinfo {title} {Fermion {{Superfluids}} of {{Nonzero Orbital Angular Momentum}} near {{Resonance}}},\ }\href {\doibase10.1103/PhysRevLett.94.090402} {\bibfield  {journal} {\bibinfo  {journal} {Phys. Rev. Lett.}\ }\textbf {\bibinfo {volume} {94}},\ \bibinfo {pages} {090402} (\bibinfo {year} {2005})}\BibitemShut {NoStop}%
\bibitem [{\citenamefont {Landau}\ and\ \citenamefont {Lifshitz}(1977)}]{landau1977quantum}%
  \BibitemOpen
  \bibfield  {author} {\bibinfo {author} {\bibfnamefont {L.~D.}\ \bibnamefont {Landau}}\ and\ \bibinfo {author} {\bibfnamefont {E.~M.}\ \bibnamefont {Lifshitz}},\ }\href@noop {} {\emph {\bibinfo {title} {Quantum Mechanics: Non-Relativistic Theory}}},\ Vol.~\bibinfo {volume} {3}\ (\bibinfo  {publisher} {{Pergamon}},\ \bibinfo {address} {Oxford},\ \bibinfo {year} {1977})\BibitemShut {NoStop}%
\bibitem [{\citenamefont {Fetter}\ and\ \citenamefont {Walecka}(2012)}]{fetter2012quantum}%
  \BibitemOpen
  \bibfield  {author} {\bibinfo {author} {\bibfnamefont {A.~L.}\ \bibnamefont {Fetter}}\ and\ \bibinfo {author} {\bibfnamefont {J.~D.}\ \bibnamefont {Walecka}},\ }\href@noop {} {\emph {\bibinfo {title} {Quantum Theory of Many-Particle Systems}}}\ (\bibinfo  {publisher} {{Courier Corporation}},\ \bibinfo {address} {New York},\ \bibinfo {year} {2012})\BibitemShut {NoStop}%
\bibitem [{\citenamefont {Ni}\ \emph {et~al.}(2010)\citenamefont {Ni}, \citenamefont {Ospelkaus}, \citenamefont {Wang}, \citenamefont {Qu{\'e}m{\'e}ner}, \citenamefont {Neyenhuis}, \citenamefont {{de Miranda}}, \citenamefont {Bohn}, \citenamefont {Ye},\ and\ \citenamefont {Jin}}]{ni2010dipolar}%
  \BibitemOpen
  \bibfield  {author} {\bibinfo {author} {\bibfnamefont {K.-K.}\ \bibnamefont {Ni}}, \bibinfo {author} {\bibfnamefont {S.}~\bibnamefont {Ospelkaus}}, \bibinfo {author} {\bibfnamefont {D.}~\bibnamefont {Wang}}, \bibinfo {author} {\bibfnamefont {G.}~\bibnamefont {Qu{\'e}m{\'e}ner}}, \bibinfo {author} {\bibfnamefont {B.}~\bibnamefont {Neyenhuis}}, \bibinfo {author} {\bibfnamefont {M.~H.~G.}\ \bibnamefont {{de Miranda}}}, \bibinfo {author} {\bibfnamefont {J.~L.}\ \bibnamefont {Bohn}}, \bibinfo {author} {\bibfnamefont {J.}~\bibnamefont {Ye}}, \ and\ \bibinfo {author} {\bibfnamefont {D.~S.}\ \bibnamefont {Jin}},\ }\bibfield  {title} {\bibinfo {title} {Dipolar collisions of polar molecules in the quantum regime},\ }\href {\doibase10.1038/nature08953} {\bibfield  {journal} {\bibinfo  {journal} {Nature (London)}\ }\textbf {\bibinfo {volume} {464}},\ \bibinfo {pages} {1324} (\bibinfo {year} {2010})}\BibitemShut {NoStop}%
\bibitem [{\citenamefont {Ospelkaus}\ \emph {et~al.}(2010)\citenamefont {Ospelkaus}, \citenamefont {Ni}, \citenamefont {Wang}, \citenamefont {De~Miranda}, \citenamefont {Neyenhuis}, \citenamefont {Qu{\'e}m{\'e}ner}, \citenamefont {Julienne}, \citenamefont {Bohn}, \citenamefont {Jin},\ and\ \citenamefont {Ye}}]{ospelkaus2010quantumstate}%
  \BibitemOpen
  \bibfield  {author} {\bibinfo {author} {\bibfnamefont {S.}~\bibnamefont {Ospelkaus}}, \bibinfo {author} {\bibfnamefont {K.-K.}\ \bibnamefont {Ni}}, \bibinfo {author} {\bibfnamefont {D.}~\bibnamefont {Wang}}, \bibinfo {author} {\bibfnamefont {M.~H.~G.}\ \bibnamefont {De~Miranda}}, \bibinfo {author} {\bibfnamefont {B.}~\bibnamefont {Neyenhuis}}, \bibinfo {author} {\bibfnamefont {.~G.}\ \bibnamefont {Qu{\'e}m{\'e}ner}}, \bibinfo {author} {\bibfnamefont {P.~S.}\ \bibnamefont {Julienne}}, \bibinfo {author} {\bibfnamefont {J.~L.}\ \bibnamefont {Bohn}}, \bibinfo {author} {\bibfnamefont {D.~S.}\ \bibnamefont {Jin}}, \ and\ \bibinfo {author} {\bibfnamefont {J.}~\bibnamefont {Ye}},\ }\bibfield  {title} {\bibinfo {title} {Quantum-state controlled chemical reactions of ultracold potassium-rubidium molecules},\ }\href {\doibase10.1126/science.1184121} {\bibfield  {journal} {\bibinfo  {journal} {Science}\ }\textbf {\bibinfo {volume} {327}},\ \bibinfo {pages} {853} (\bibinfo {year} {2010})}\BibitemShut {NoStop}%
\bibitem [{\citenamefont {De~Marco}\ \emph {et~al.}(2019)\citenamefont {De~Marco}, \citenamefont {Valtolina}, \citenamefont {Matsuda}, \citenamefont {Tobias}, \citenamefont {Covey},\ and\ \citenamefont {Ye}}]{demarco2019degenerate}%
  \BibitemOpen
  \bibfield  {author} {\bibinfo {author} {\bibfnamefont {L.}~\bibnamefont {De~Marco}}, \bibinfo {author} {\bibfnamefont {G.}~\bibnamefont {Valtolina}}, \bibinfo {author} {\bibfnamefont {K.}~\bibnamefont {Matsuda}}, \bibinfo {author} {\bibfnamefont {W.~G.}\ \bibnamefont {Tobias}}, \bibinfo {author} {\bibfnamefont {J.~P.}\ \bibnamefont {Covey}}, \ and\ \bibinfo {author} {\bibfnamefont {J.}~\bibnamefont {Ye}},\ }\bibfield  {title} {\bibinfo {title} {A degenerate {{Fermi}} gas of polar molecules},\ }\href {\doibase10.1126/science.aau7230} {\bibfield  {journal} {\bibinfo  {journal} {Science}\ }\textbf {\bibinfo {volume} {363}},\ \bibinfo {pages} {853} (\bibinfo {year} {2019})}\BibitemShut {NoStop}%
\bibitem [{\citenamefont {Liu}(2013)}]{liu2013virial}%
  \BibitemOpen
  \bibfield  {author} {\bibinfo {author} {\bibfnamefont {X.-J.}\ \bibnamefont {Liu}},\ }\bibfield  {title} {\bibinfo {title} {Virial expansion for a strongly correlated {{Fermi}} system and its application to ultracold atomic {{Fermi}} gases},\ }\href {\doibase10.1016/j.physrep.2012.10.004} {\bibfield  {journal} {\bibinfo  {journal} {Phys. Rep.}\ }\textbf {\bibinfo {volume} {524}},\ \bibinfo {pages} {37} (\bibinfo {year} {2013})}\BibitemShut {NoStop}%
\bibitem [{\citenamefont {Marcelino}\ \emph {et~al.}(2014)\citenamefont {Marcelino}, \citenamefont {Nicolai}, \citenamefont {Roditi},\ and\ \citenamefont {LeClair}}]{marcelino2014virial}%
  \BibitemOpen
  \bibfield  {author} {\bibinfo {author} {\bibfnamefont {E.}~\bibnamefont {Marcelino}}, \bibinfo {author} {\bibfnamefont {A.}~\bibnamefont {Nicolai}}, \bibinfo {author} {\bibfnamefont {I.}~\bibnamefont {Roditi}}, \ and\ \bibinfo {author} {\bibfnamefont {A.}~\bibnamefont {LeClair}},\ }\bibfield  {title} {\bibinfo {title} {Virial coefficients for trapped {{Bose}} and {{Fermi}} gases beyond the unitary limit: {{An}} {$S$}-matrix approach},\ }\href {\doibase10.1103/PhysRevA.90.053619} {\bibfield  {journal} {\bibinfo  {journal} {Phys. Rev. A}\ }\textbf {\bibinfo {volume} {90}},\ \bibinfo {pages} {053619} (\bibinfo {year} {2014})}\BibitemShut {NoStop}%
\bibitem [{\citenamefont {Dalfovo}\ \emph {et~al.}(1999)\citenamefont {Dalfovo}, \citenamefont {Giorgini}, \citenamefont {Pitaevskii},\ and\ \citenamefont {Stringari}}]{dalfovo1999theory}%
  \BibitemOpen
  \bibfield  {author} {\bibinfo {author} {\bibfnamefont {F.}~\bibnamefont {Dalfovo}}, \bibinfo {author} {\bibfnamefont {S.}~\bibnamefont {Giorgini}}, \bibinfo {author} {\bibfnamefont {L.~P.}\ \bibnamefont {Pitaevskii}}, \ and\ \bibinfo {author} {\bibfnamefont {S.}~\bibnamefont {Stringari}},\ }\bibfield  {title} {\bibinfo {title} {Theory of {{Bose-Einstein}} condensation in trapped gases},\ }\href {\doibase10.1103/RevModPhys.71.463} {\bibfield  {journal} {\bibinfo  {journal} {Rev. Mod. Phys.}\ }\textbf {\bibinfo {volume} {71}},\ \bibinfo {pages} {463} (\bibinfo {year} {1999})}\BibitemShut {NoStop}%
\bibitem [{\citenamefont {Butts}\ and\ \citenamefont {Rokhsar}(1997)}]{butts1997trapped}%
  \BibitemOpen
  \bibfield  {author} {\bibinfo {author} {\bibfnamefont {D.~A.}\ \bibnamefont {Butts}}\ and\ \bibinfo {author} {\bibfnamefont {D.~S.}\ \bibnamefont {Rokhsar}},\ }\bibfield  {title} {\bibinfo {title} {Trapped fermi gases},\ }\href {\doibase10.1103/PhysRevA.55.4346} {\bibfield  {journal} {\bibinfo  {journal} {Phys. Rev. A}\ }\textbf {\bibinfo {volume} {55}},\ \bibinfo {pages} {4346} (\bibinfo {year} {1997})}\BibitemShut {NoStop}%
\bibitem [{\citenamefont {Bethe}(1935)}]{bethe1935theory}%
  \BibitemOpen
  \bibfield  {author} {\bibinfo {author} {\bibfnamefont {H.~A.}\ \bibnamefont {Bethe}},\ }\bibfield  {title} {\bibinfo {title} {Theory of disintegration of nuclei by neutrons},\ }\href {\doibase10.1103/PhysRev.47.747} {\bibfield  {journal} {\bibinfo  {journal} {Phys. Rev.}\ }\textbf {\bibinfo {volume} {47}},\ \bibinfo {pages} {747} (\bibinfo {year} {1935})}\BibitemShut {NoStop}%
\bibitem [{\citenamefont {Wigner}(1948)}]{wigner1948behavior}%
  \BibitemOpen
  \bibfield  {author} {\bibinfo {author} {\bibfnamefont {E.~P.}\ \bibnamefont {Wigner}},\ }\bibfield  {title} {\bibinfo {title} {On the behavior of cross sections near thresholds},\ }\href {\doibase10.1103/PhysRev.73.1002} {\bibfield  {journal} {\bibinfo  {journal} {Phys. Rev.}\ }\textbf {\bibinfo {volume} {73}},\ \bibinfo {pages} {1002} (\bibinfo {year} {1948})}\BibitemShut {NoStop}%
\bibitem [{\citenamefont {Sadeghpour}\ \emph {et~al.}(2000)\citenamefont {Sadeghpour}, \citenamefont {Bohn}, \citenamefont {Cavagnero}, \citenamefont {Esry}, \citenamefont {Fabrikant}, \citenamefont {Macek},\ and\ \citenamefont {Rau}}]{sadeghpour2000collisions}%
  \BibitemOpen
  \bibfield  {author} {\bibinfo {author} {\bibfnamefont {H.~R.}\ \bibnamefont {Sadeghpour}}, \bibinfo {author} {\bibfnamefont {J.~L.}\ \bibnamefont {Bohn}}, \bibinfo {author} {\bibfnamefont {M.~J.}\ \bibnamefont {Cavagnero}}, \bibinfo {author} {\bibfnamefont {B.~D.}\ \bibnamefont {Esry}}, \bibinfo {author} {\bibfnamefont {I.~I.}\ \bibnamefont {Fabrikant}}, \bibinfo {author} {\bibfnamefont {J.~H.}\ \bibnamefont {Macek}}, \ and\ \bibinfo {author} {\bibfnamefont {A.~R.~P.}\ \bibnamefont {Rau}},\ }\bibfield  {title} {\bibinfo {title} {Collisions near threshold in atomic and molecular physics},\ }\href {\doibase10.1088/0953-4075/33/5/201} {\bibfield  {journal} {\bibinfo  {journal} {J. Phys. B}\ }\textbf {\bibinfo {volume} {33}},\ \bibinfo {pages} {R93} (\bibinfo {year} {2000})}\BibitemShut {NoStop}%
\bibitem [{\citenamefont {Quéméner}\ and\ \citenamefont {Bohn}(2010)}]{quemener2010stronga}%
  \BibitemOpen
  \bibfield  {author} {\bibinfo {author} {\bibfnamefont {G.}~\bibnamefont {Quéméner}}\ and\ \bibinfo {author} {\bibfnamefont {J.~L.}\ \bibnamefont {Bohn}},\ }\bibfield  {title} {\bibinfo {title} {Strong dependence of ultracold chemical rates on electric dipole moments},\ }\href {\doibase10.1103/PhysRevA.81.022702} {\bibfield  {journal} {\bibinfo  {journal} {Phys. Rev. A}\ }\textbf {\bibinfo {volume} {81}},\ \bibinfo {pages} {022702} (\bibinfo {year} {2010})}\BibitemShut {NoStop}%
\bibitem [{\citenamefont {Jachymski}\ \emph {et~al.}(2014)\citenamefont {Jachymski}, \citenamefont {Krych}, \citenamefont {Julienne},\ and\ \citenamefont {Idziaszek}}]{jachymski2014quantumdefect}%
  \BibitemOpen
  \bibfield  {author} {\bibinfo {author} {\bibfnamefont {K.}~\bibnamefont {Jachymski}}, \bibinfo {author} {\bibfnamefont {M.}~\bibnamefont {Krych}}, \bibinfo {author} {\bibfnamefont {P.~S.}\ \bibnamefont {Julienne}}, \ and\ \bibinfo {author} {\bibfnamefont {Z.}~\bibnamefont {Idziaszek}},\ }\bibfield  {title} {\bibinfo {title} {Quantum-defect model of a reactive collision at finite temperature},\ }\href {\doibase10.1103/PhysRevA.90.042705} {\bibfield  {journal} {\bibinfo  {journal} {Phys. Rev. A}\ }\textbf {\bibinfo {volume} {90}},\ \bibinfo {pages} {042705} (\bibinfo {year} {2014})}\BibitemShut {NoStop}%
\bibitem [{\citenamefont {Waseem}\ \emph {et~al.}(2017)\citenamefont {Waseem}, \citenamefont {Saito}, \citenamefont {Yoshida},\ and\ \citenamefont {Mukaiyama}}]{waseem2017twobody}%
  \BibitemOpen
  \bibfield  {author} {\bibinfo {author} {\bibfnamefont {M.}~\bibnamefont {Waseem}}, \bibinfo {author} {\bibfnamefont {T.}~\bibnamefont {Saito}}, \bibinfo {author} {\bibfnamefont {J.}~\bibnamefont {Yoshida}}, \ and\ \bibinfo {author} {\bibfnamefont {T.}~\bibnamefont {Mukaiyama}},\ }\bibfield  {title} {\bibinfo {title} {Two-body relaxation in a {{Fermi}} gas at a $p$-wave {{Feshbach}} resonance},\ }\href {\doibase10.1103/PhysRevA.96.062704} {\bibfield  {journal} {\bibinfo  {journal} {Phys. Rev. A}\ }\textbf {\bibinfo {volume} {96}},\ \bibinfo {pages} {062704} (\bibinfo {year} {2017})}\BibitemShut {NoStop}%
\bibitem [{\citenamefont {Braaten}\ \emph {et~al.}(2008)\citenamefont {Braaten}, \citenamefont {Hammer}, \citenamefont {Kang},\ and\ \citenamefont {Platter}}]{braaten2008threebody}%
  \BibitemOpen
  \bibfield  {author} {\bibinfo {author} {\bibfnamefont {E.}~\bibnamefont {Braaten}}, \bibinfo {author} {\bibfnamefont {H.-W.}\ \bibnamefont {Hammer}}, \bibinfo {author} {\bibfnamefont {D.}~\bibnamefont {Kang}}, \ and\ \bibinfo {author} {\bibfnamefont {L.}~\bibnamefont {Platter}},\ }\bibfield  {title} {\bibinfo {title} {Three-body recombination of identical bosons with a large positive scattering length at nonzero temperature},\ }\href {\doibase10.1103/PhysRevA.78.043605} {\bibfield  {journal} {\bibinfo  {journal} {Phys. Rev. A}\ }\textbf {\bibinfo {volume} {78}},\ \bibinfo {pages} {043605} (\bibinfo {year} {2008})}\BibitemShut {NoStop}%
\bibitem [{\citenamefont {Pethick}\ and\ \citenamefont {Smith}(2008)}]{pethick2008bose}%
  \BibitemOpen
  \bibfield  {author} {\bibinfo {author} {\bibfnamefont {C.~J.}\ \bibnamefont {Pethick}}\ and\ \bibinfo {author} {\bibfnamefont {H.}~\bibnamefont {Smith}},\ }\href@noop {} {\emph {\bibinfo {title} {Bose\textendash{{Einstein}} Condensation in Dilute Gases}}}\ (\bibinfo  {publisher} {{Cambridge University Press}},\ \bibinfo {address} {Cambridge},\ \bibinfo {year} {2008})\BibitemShut {NoStop}%
\bibitem [{\citenamefont {Gregory}\ \emph {et~al.}(2019)\citenamefont {Gregory}, \citenamefont {Frye}, \citenamefont {Blackmore}, \citenamefont {Bridge}, \citenamefont {Sawant}, \citenamefont {Hutson},\ and\ \citenamefont {Cornish}}]{gregory2019sticky}%
  \BibitemOpen
  \bibfield  {author} {\bibinfo {author} {\bibfnamefont {P.~D.}\ \bibnamefont {Gregory}}, \bibinfo {author} {\bibfnamefont {M.~D.}\ \bibnamefont {Frye}}, \bibinfo {author} {\bibfnamefont {J.~A.}\ \bibnamefont {Blackmore}}, \bibinfo {author} {\bibfnamefont {E.~M.}\ \bibnamefont {Bridge}}, \bibinfo {author} {\bibfnamefont {R.}~\bibnamefont {Sawant}}, \bibinfo {author} {\bibfnamefont {J.~M.}\ \bibnamefont {Hutson}}, \ and\ \bibinfo {author} {\bibfnamefont {S.~L.}\ \bibnamefont {Cornish}},\ }\bibfield  {title} {\bibinfo {title} {Sticky collisions of ultracold {{RbCs}} molecules},\ }\href {\doibase10.1038/s41467-019-11033-y} {\bibfield  {journal} {\bibinfo  {journal} {Nat. Commun.}\ }\textbf {\bibinfo {volume} {10}},\ \bibinfo {pages} {3104} (\bibinfo {year} {2019})}\BibitemShut {NoStop}%
\bibitem [{\citenamefont {Idziaszek}\ and\ \citenamefont {Julienne}(2010)}]{idziaszek2010universal}%
  \BibitemOpen
  \bibfield  {author} {\bibinfo {author} {\bibfnamefont {Z.}~\bibnamefont {Idziaszek}}\ and\ \bibinfo {author} {\bibfnamefont {P.~S.}\ \bibnamefont {Julienne}},\ }\bibfield  {title} {\bibinfo {title} {Universal rate constants for reactive collisions of ultracold molecules},\ }\href {\doibase10.1103/PhysRevLett.104.113202} {\bibfield  {journal} {\bibinfo  {journal} {Phys. Rev. Lett.}\ }\textbf {\bibinfo {volume} {104}},\ \bibinfo {pages} {113202} (\bibinfo {year} {2010})}\BibitemShut {NoStop}%
\bibitem [{\citenamefont {Venu}\ \emph {et~al.}(2023)\citenamefont {Venu}, \citenamefont {Xu}, \citenamefont {Mamaev}, \citenamefont {Corapi}, \citenamefont {Bilitewski}, \citenamefont {D'Incao}, \citenamefont {Fujiwara}, \citenamefont {Rey},\ and\ \citenamefont {Thywissen}}]{venu2023unitarya}%
  \BibitemOpen
  \bibfield  {author} {\bibinfo {author} {\bibfnamefont {V.}~\bibnamefont {Venu}}, \bibinfo {author} {\bibfnamefont {P.}~\bibnamefont {Xu}}, \bibinfo {author} {\bibfnamefont {M.}~\bibnamefont {Mamaev}}, \bibinfo {author} {\bibfnamefont {F.}~\bibnamefont {Corapi}}, \bibinfo {author} {\bibfnamefont {T.}~\bibnamefont {Bilitewski}}, \bibinfo {author} {\bibfnamefont {J.~P.}\ \bibnamefont {D'Incao}}, \bibinfo {author} {\bibfnamefont {C.~J.}\ \bibnamefont {Fujiwara}}, \bibinfo {author} {\bibfnamefont {A.~M.}\ \bibnamefont {Rey}}, \ and\ \bibinfo {author} {\bibfnamefont {J.~H.}\ \bibnamefont {Thywissen}},\ }\bibfield  {title} {\bibinfo {title} {{Unitary $p$-Wave Interactions between Fermions in an Optical Lattice}},\ }\href {\doibase10.1038/s41586-022-05405-6} {\bibfield  {journal} {\bibinfo  {journal} {Nature (London)}\ }\textbf {\bibinfo {volume} {613}},\ \bibinfo {pages} {262} (\bibinfo {year} {2023})}\BibitemShut {NoStop}%
\bibitem [{\citenamefont {Gerken}\ \emph {et~al.}(2019)\citenamefont {Gerken}, \citenamefont {Tran}, \citenamefont {H{\"a}fner}, \citenamefont {Tiemann}, \citenamefont {Zhu},\ and\ \citenamefont {Weidem{\"u}ller}}]{gerken2019observation}%
  \BibitemOpen
  \bibfield  {author} {\bibinfo {author} {\bibfnamefont {M.}~\bibnamefont {Gerken}}, \bibinfo {author} {\bibfnamefont {B.}~\bibnamefont {Tran}}, \bibinfo {author} {\bibfnamefont {S.}~\bibnamefont {H{\"a}fner}}, \bibinfo {author} {\bibfnamefont {E.}~\bibnamefont {Tiemann}}, \bibinfo {author} {\bibfnamefont {B.}~\bibnamefont {Zhu}}, \ and\ \bibinfo {author} {\bibfnamefont {M.}~\bibnamefont {Weidem{\"u}ller}},\ }\bibfield  {title} {\bibinfo {title} {{Observation of Dipolar Splittings in High-Resolution Atom-Loss Spectroscopy of ${}^{6}\mathrm{Li}$ $p$-Wave Feshbach Resonances}},\ }\href {\doibase10.1103/PhysRevA.100.050701} {\bibfield  {journal} {\bibinfo  {journal} {Phys. Rev. A}\ }\textbf {\bibinfo {volume} {100}},\ \bibinfo {pages} {050701} (\bibinfo {year} {2019})}\BibitemShut {NoStop}%
\bibitem [{\citenamefont {Sakurai}\ and\ \citenamefont {Commins}(1994)}]{sakurai1994modern}%
  \BibitemOpen
  \bibfield  {author} {\bibinfo {author} {\bibfnamefont {J.~J.}\ \bibnamefont {Sakurai}}\ and\ \bibinfo {author} {\bibfnamefont {E.~D.}\ \bibnamefont {Commins}},\ }\href@noop {} {\emph {\bibinfo {title} {Modern Quantum Mechanics, Revised Edition}}}\ (\bibinfo  {publisher} {{Addison-Wesley Publishing Company}},\ \bibinfo {address} {Reading, MA},\ \bibinfo {year} {1994})\BibitemShut {NoStop}%
\bibitem [{\citenamefont {Zhai}(2021)}]{zhai2021ultracold}%
  \BibitemOpen
  \bibfield  {author} {\bibinfo {author} {\bibfnamefont {H.}~\bibnamefont {Zhai}},\ }\href@noop {} {\emph {\bibinfo {title} {Ultracold {{Atomic Physics}}}}}\ (\bibinfo  {publisher} {{Cambridge University Press}},\ \bibinfo {address} {Cambridge},\ \bibinfo {year} {2021})\BibitemShut {NoStop}%
\bibitem [{\citenamefont {Braaten}\ and\ \citenamefont {Hammer}(2007)}]{braaten2007efimov}%
  \BibitemOpen
  \bibfield  {author} {\bibinfo {author} {\bibfnamefont {E.}~\bibnamefont {Braaten}}\ and\ \bibinfo {author} {\bibfnamefont {H.~W.}\ \bibnamefont {Hammer}},\ }\bibfield  {title} {\bibinfo {title} {{Efimov Physics in Cold Atoms}},\ }\href {\doibase10.1016/j.aop.2006.10.011} {\bibfield  {journal} {\bibinfo  {journal} {Ann. Phys.}\ }\textbf {\bibinfo {volume} {322}},\ \bibinfo {pages} {120} (\bibinfo {year} {2007})}\BibitemShut {NoStop}%
\bibitem [{\citenamefont {Weisstein}()}]{weissteintrinomial}%
  \BibitemOpen
  \bibfield  {author} {\bibinfo {author} {\bibfnamefont {E.~W.}\ \bibnamefont {Weisstein}},\ }\href@noop {} {\bibinfo {title} {Trinomial {{Coefficient}}},\ }\bibinfo {howpublished} {https://mathworld.wolfram.com/}\BibitemShut {NoStop}%
\bibitem [{\citenamefont {Mott}\ and\ \citenamefont {Massey}(1971)}]{mott1971theory}%
  \BibitemOpen
  \bibfield  {author} {\bibinfo {author} {\bibfnamefont {N.~F.}\ \bibnamefont {Mott}}\ and\ \bibinfo {author} {\bibfnamefont {H.~S.~W.}\ \bibnamefont {Massey}},\ }\href@noop {} {\emph {\bibinfo {title} {Theory of {{Atomic Collisions}}}}}\ (\bibinfo  {publisher} {{Oxford University Press}},\ \bibinfo {address} {{Oxford}},\ \bibinfo {year} {1971})\BibitemShut {NoStop}%
\bibitem [{\citenamefont {Balakrishnan}\ \emph {et~al.}(1997)\citenamefont {Balakrishnan}, \citenamefont {Kharchenko}, \citenamefont {Forrey},\ and\ \citenamefont {Dalgarno}}]{balakrishnan1997complex}%
  \BibitemOpen
  \bibfield  {author} {\bibinfo {author} {\bibfnamefont {N.}~\bibnamefont {Balakrishnan}}, \bibinfo {author} {\bibfnamefont {V.}~\bibnamefont {Kharchenko}}, \bibinfo {author} {\bibfnamefont {R.~C.}\ \bibnamefont {Forrey}}, \ and\ \bibinfo {author} {\bibfnamefont {A.}~\bibnamefont {Dalgarno}},\ }\bibfield  {title} {\bibinfo {title} {Complex scattering lengths in multi-channel atom{\textendash}molecule collisions},\ }\href {\doibase10.1016/S0009-2614(97)01052-X} {\bibfield  {journal} {\bibinfo  {journal} {Chem. Phys. Lett.}\ }\textbf {\bibinfo {volume} {280}},\ \bibinfo {pages} {5} (\bibinfo {year} {1997})}\BibitemShut {NoStop}%
\bibitem [{\citenamefont {Kurlov}\ and\ \citenamefont {Shlyapnikov}(2017)}]{kurlov2017twobody}%
  \BibitemOpen
  \bibfield  {author} {\bibinfo {author} {\bibfnamefont {D.~V.}\ \bibnamefont {Kurlov}}\ and\ \bibinfo {author} {\bibfnamefont {G.~V.}\ \bibnamefont {Shlyapnikov}},\ }\bibfield  {title} {\bibinfo {title} {Two-body relaxation of spin-polarized fermions in reduced dimensionalities near a $p$-wave {{Feshbach}} resonance},\ }\href {\doibase10.1103/PhysRevA.95.032710} {\bibfield  {journal} {\bibinfo  {journal} {Phys. Rev. A}\ }\textbf {\bibinfo {volume} {95}},\ \bibinfo {pages} {032710} (\bibinfo {year} {2017})}\BibitemShut {NoStop}%
\bibitem [{\citenamefont {Croft}\ \emph {et~al.}(2020)\citenamefont {Croft}, \citenamefont {Bohn},\ and\ \citenamefont {Qu{\'e}m{\'e}ner}}]{croft2020unified}%
  \BibitemOpen
  \bibfield  {author} {\bibinfo {author} {\bibfnamefont {J.~F.~E.}\ \bibnamefont {Croft}}, \bibinfo {author} {\bibfnamefont {J.~L.}\ \bibnamefont {Bohn}}, \ and\ \bibinfo {author} {\bibfnamefont {G.}~\bibnamefont {Qu{\'e}m{\'e}ner}},\ }\bibfield  {title} {\bibinfo {title} {Unified model of ultracold molecular collisions},\ }\href {\doibase10.1103/PhysRevA.102.033306} {\bibfield  {journal} {\bibinfo  {journal} {Phys. Rev. A}\ }\textbf {\bibinfo {volume} {102}},\ \bibinfo {pages} {033306} (\bibinfo {year} {2020})}\BibitemShut {NoStop}%
\bibitem [{\citenamefont {Abramowitz}\ \emph {et~al.}(1988)\citenamefont {Abramowitz}, \citenamefont {Stegun},\ and\ \citenamefont {Romer}}]{abramowitz1988handbook}%
  \BibitemOpen
  \bibfield  {author} {\bibinfo {author} {\bibfnamefont {M.}~\bibnamefont {Abramowitz}}, \bibinfo {author} {\bibfnamefont {I.~A.}\ \bibnamefont {Stegun}}, \ and\ \bibinfo {author} {\bibfnamefont {R.~H.}\ \bibnamefont {Romer}},\ }\href@noop {} {\emph {\bibinfo {title} {{Handbook of Mathematical Functions with Formulas, Graphs, and Mathematical Tables}}}}\ (\bibinfo  {publisher} {Dover},\ \bibinfo {address} {New York},\ \bibinfo {year} {1988})\BibitemShut {NoStop}%
\bibitem [{\citenamefont {Wood}(1992)}]{wood1992computation}%
  \BibitemOpen
  \bibfield  {author} {\bibinfo {author} {\bibfnamefont {D.}~\bibnamefont {Wood}},\ }\href@noop {} {\emph {\bibinfo {title} {The Computation of Polylogarithms}}},\ \bibinfo {type} {Tech. Rep.}\ \bibinfo {number} {15-92*}\ (\bibinfo  {institution} {{University of Kent, Computing Laboratory}},\ \bibinfo {address} {{University of Kent, Canterbury, UK}},\ \bibinfo {year} {1992})\BibitemShut {NoStop}%
\end{thebibliography}
\end{document}